\documentclass{article}

\usepackage{arxiv}

\usepackage[utf8]{inputenc} 
\usepackage[T1]{fontenc}    
\usepackage{url}            
\usepackage{booktabs}       
\usepackage{amsfonts}       
\usepackage{nicefrac}       
\usepackage{microtype}      
\usepackage{lipsum}		
\usepackage{graphicx}
\usepackage{doi}
\usepackage{subcaption}
\usepackage[justification=centering]{caption}
\usepackage{amsmath}
\usepackage{graphicx}

\title{DODEM: \underline{DO}uble \underline{DE}fense \underline{M}echanism Against Adversarial Attacks Towards Secure Industrial Internet of Things Analytics}

\author{{\hspace{1mm}Onat Gungor} \\
	Department of Electrical and Computer Engineering\\
	University of California, San Diego\\
	\texttt{ogungor@ucsd.edu} \\
	\AND
	Tajana Rosing \\
	Department of Electrical and Computer Engineering\\
	University of California, San Diego \\
	\texttt{tajana@ucsd.edu} \\
	\AND
	Baris Aksanli \\
	Department of Electrical and Computer Engineering \\
	San Diego State University \\
	\texttt{baksanli@sdsu.edu} \\
}

\renewcommand{\shorttitle}{DODEM: Double Defense Mechanism Against Adversarial Attacks}

\begin{document}

\maketitle
\begin{abstract}
Industrial Internet of Things (I-IoT) is a collaboration of devices, sensors, and networking equipment to monitor and collect data from industrial operations. Machine learning (ML) methods use this data to make high-level decisions with minimal human intervention. Data-driven predictive maintenance (PDM) is a crucial ML-based I-IoT application to find an optimal maintenance schedule for industrial assets. The performance of these ML methods can seriously be threatened by adversarial attacks where an adversary crafts perturbed data and sends it to the ML model to deteriorate its prediction performance. The models should be able to stay robust against these attacks where robustness is measured by how much perturbation in input data affects model performance. Hence, there is a need for effective defense mechanisms that can protect these models against adversarial attacks. In this work, we propose a double defense mechanism to detect and mitigate adversarial attacks in I-IoT environments. We first detect if there is an adversarial attack on a given sample using novelty detection algorithms. Then, based on the outcome of our algorithm, marking an instance as attack or normal, we select adversarial retraining or standard training to provide a secondary defense layer. If there is an attack, adversarial retraining provides a more robust model, while we apply standard training for regular samples. Since we may not know if an attack will take place, our adaptive mechanism allows us to consider irregular changes in data. The results show that our double defense strategy is highly efficient where we can improve model robustness by up to 64.6\% and 52\% compared to standard and adversarial retraining, respectively. 
\end{abstract}


\section{Introduction}
\begin{figure}[t]
\centering
\includegraphics[width=0.95\linewidth]
{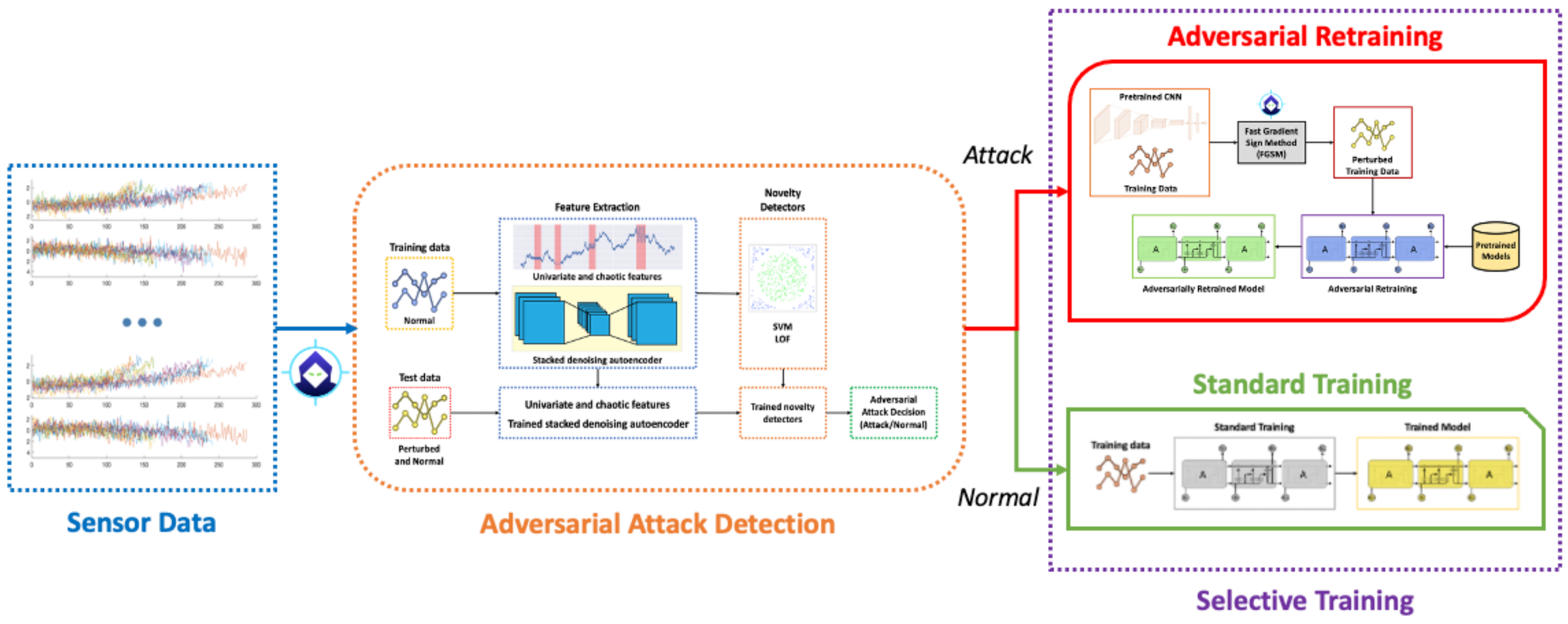}
\caption{Our Proposed Framework (\textit{DODEM})}
\label{framework}
\end{figure}

Cyber-physical systems (CPSs) have become extremely intelligent and autonomous due to significant technological advancements in the Internet of Things (IoT) and machine learning (ML) \cite{li2020adversarial}. Learning algorithms utilize collected sensor data to  provide insight into high-level business decisions. Industrial Internet of Things (I-IoT) is a typical CPS enabling the operation, interconnection, and intelligence of industrial systems \cite{xu2018survey}. I-IoT's small-scale devices, with their limited computation and communication capabilities, make them vulnerable to potential cyber attacks \cite{gungor2022stewart}. An adversary can exploit these vulnerabilities to sabotage communication, prevent asset availability, and corrupt monitoring data which may have serious financial consequences \cite{tuptuk2018security}. For instance, the average estimated losses were \$10.7 million per breach of data among manufacturing organizations in Asia Pacific in 2019 \cite{microsoft2019}. Among different malicious attacks, adversarial attacks against deep learning (DL) draw considerable attention as DL methods are becoming more widely adopted. In an adversarial attack, an adversary can access the DL model to create slight but carefully-crafted perturbed examples to deteriorate the model performance. These attacks are especially a grave threat to data-driven predictive maintenance which aims to find an optimal maintenance schedule based on time-to-failure prediction of an industrial asset. Data-driven remaining useful life (RUL) estimation is an essential task in predictive maintenance, and it became popular owing to abundance of I-IoT system monitoring data \cite{gungor2021dowell}. It uses ML methods to map sensor input to the corresponding RUL values. Since ML is in the center of data-driven RUL prediction, adversarial attacks may have serious consequences such as wrong maintenance decisions leading to undetected failures in a system \cite{mode2020crafting}. Hence, there is a need for detecting and mitigating such attacks to maintain the integrity and correct functioning of I-IoT systems \cite{abdu2020detecting}. 

Adversarial attack detection aims to distinguish adversarial samples from legitimate ones \cite{metzen2017detecting}. Detecting adversarial inputs can provide an additional protection when other defense lines have already been breached \cite{santana2021detecting}. Supervised or semi-supervised ML methods can be adapted for adversarial attack detection \cite{yan2019attack}. Semi-supervised training presents a more realistic learning process since training data may not contain any adversarial samples. For semi-supervised attack detection, training starts with clean data feature extraction. Those features are then provided to ML model for its training. In testing, adversarial examples are injected into clean test data as a first step. After test data features are extracted, trained ML model makes a binary decision representing if there is an attack or not. This task is defined as \textit{novelty detection} which classifies test data that differs in some way from the data that are available during training \cite{pimentel2014review}. Novelty detection is especially suitable for adversarial attack detection because usually the amount of adversarial examples is extremely small compared to legitimate ones. There are various novelty detection algorithms proposed in the literature such as Bayesian approaches, extreme value statistics, support vector methods, and neural networks \cite{pimentel2014review}. Adversarial examples are not easy to detect \cite{carlini2017adversarial} and the extracted feature quality can affect the prediction performance significantly. Hence, there is a need for additional defense layers that can minimize the impact of wrong adversarial attack detection decisions.    

Adversarial (re)training is one of the most effective approaches to defend DL models against adversarial attacks \cite{bai2021recent}. It augments the original training data with adversarial examples at each iteration, leading to more robust models when facing adversarial attacks \cite{madry2017towards}. The quality of adversarial examples included in adversarial (re)training plays a crucial role in model robustness. To craft those instances, we can adopt numerous real-world attack patterns \cite{warr2019strengthening}: direct attack, replica attack, and transfer attack. While the first two have some information about the target DNN (e.g., submitting inputs to the target and receiving results, using an exact replica of the target model, etc.), transfer attack does not assume anything about the model, and thus, it represents a more feasible approach in many real-life attack scenarios. In a transfer attack, an attacker can use a good-enough approximation (i.e., substitute model) of the target model to craft adversarial inputs. Even though a substitute model is different from the target model, \textit{transferability property} makes transfer attacks extremely dangerous. This property is satisfied when an attack developed for a substitute model is also effective against the target model \cite{demontis2019adversarial}. It has been shown that two models with different architectures are likely to be susceptible to similar adversarial examples if they have been trained with the same data \cite{warr2019strengthening}. However, this assumption may not hold if an adversary does not have access to the same data for substitute and target models. Hence, adversarial (re)training should be modified by reconsidering this assumption. Furthermore, another problem with adversarial (re)training is its generalization ability. Although adversarial (re)training increases the robustness on adversarial examples, it might negatively affect the clean data accuracy, i.e., models fortified with adversarial (re)training performs worse when there is no attack\cite{raghunathan2019adversarial}. This necessitates an alternative solution when there is no attack. 

In this work, we propose a DOuble DEfense Mechanism framework (DODEM) that combines adversarial attack detection with another defense layer, selective training, as illustrated in Figure \ref{framework}. Given sensor data, our adversarial attack detection module first determines if a sample is adversarial or not. Based on this input, the second defense module, selective training, chooses adversarial retraining or standard training. If there is an attack, our adversarial retraining approach is employed as the attack mitigation strategy to strengthen model robustness on adversarial data. If the attack detection algorithm does not detect any attack, we choose standard training strategy due to its high accuracy on clean data. For adversarial attack detection, we extract features based on multiple techniques, including statistical features, chaos-theoretic measures, and stacked denoising auto encoder. We then train two novelty detection algorithms: one-class support vector machine (OCSVM) \cite{cortes1995support} and local outlier factor (LOF) \cite{breunig2000lof}. The results show that LOF predictions are significantly better than OCSVM where LOF has 89\% average $F_{2}$ score while OCSVM can reach 63\%. For selective training, we have two options: adversarial retraining or standard training. For adversarial retraining, we first observe that a substitute model trained on a different dataset can fool the target model more compared to training with the same dataset. Based on this observation, we modify traditional adversarial (re)training process. We first train our substitute and target models on different datasets. We then transfer the trained substitute model to create perturbed training examples and include them in adversarial (re)training process of the target model. For standard training, we use the original (clean) training dataset. Overall, our proposed solution presents a two-layer defense mechanism against adversarial attacks. Thus, it becomes more difficult for an adversarial attack to deteriorate the ML prediction performance. We test our defense strategy on the NASA Commercial Modular Aero-Propulsion System Simulation (C-MAPSS) data set \cite{saxena2008damage} which is a commonly-used benchmark dataset for RUL prediction. We compare the robustness of our strategy with standard training, and our adversarial retraining approach under white-box adversarial attacks. The results show that our method consistently outperforms standard training and adversarial retraining by adaptively selecting the correct defense strategy. Our proposed defense is by up to 64.6\% and 52\% more robust compared to standard and adversarial retraining respectively under different adversarial attacks. To the
best of our knowledge, our work is the first that proposes a selective defense mechanism towards more robust data-driven
predictive maintenance. We can summarize our contributions as follows:
\begin{itemize}
    \item accurate adversarial attack detection approach built on state-of-the-art feature extraction and novelty detection methods,  
    \item novel and realistic transfer attack-based adversarial retraining strategy which uses models trained on different datasets to create perturbed training instances, 
    \item selective training approach as an additional defense layer that can decide adversarial retraining or standard training based on adversarial attack detection algorithm output,
    \item effective defense mechanism that can minimize the impact of adversarial attacks  under white-box attacks compared to standard training and adversarial retraining.
\end{itemize}


\section{Related Work}
\begin{figure}[t]
\centering
\includegraphics[width=0.9\linewidth]
{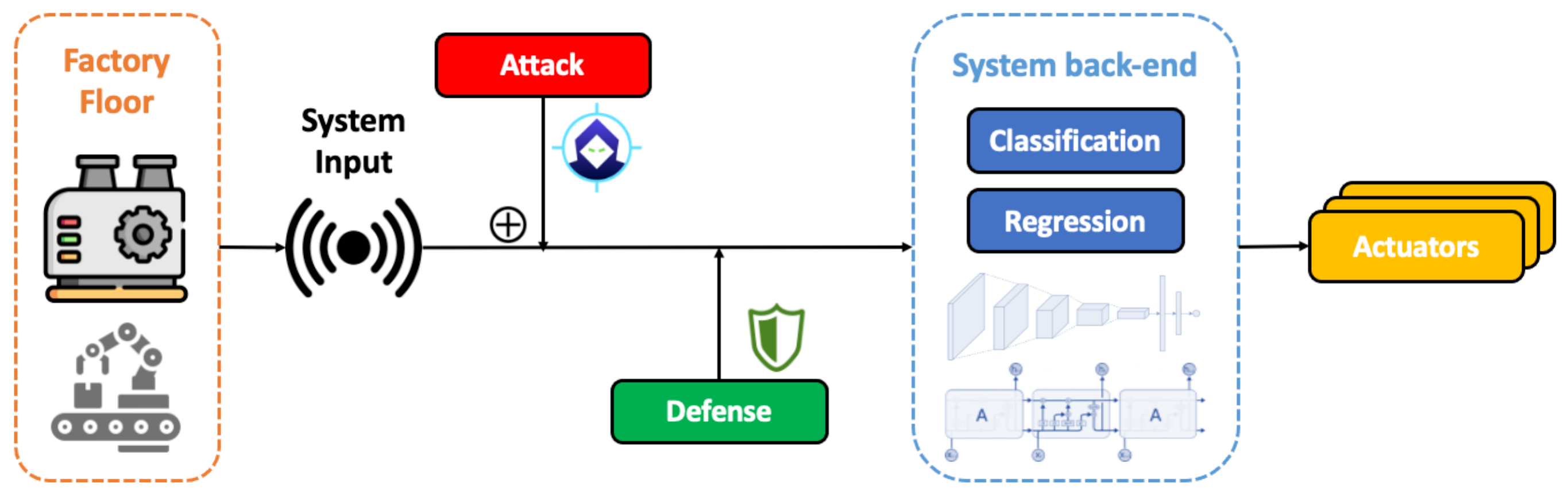}
\caption{I-IoT Adversarial Attack Workflow}
\label{attack-workflow}
\end{figure}

I-IoT brings full automation, higher reliability, and fine-grained control to the manufacturing world, utilizing computer networks to collect data from connected machines, and convert this data into actionable information \cite{xu2018survey}. However, these systems are susceptible to cyber attacks due to inadequate standardization, and the lack of required skills to implement them \cite{lezzi2018cybersecurity}. To cope with this challenge, cyber security measures should be taken such as cyber-security awareness training, keeping software programs up-to-date, installing a firewall, and using strong passwords \cite{he2019improving}. There are several different cyber attacks that can target I-IoT systems, e.g., denial of service, eavesdropping, man-in-the-middle, side channel, zero day, and adversarial attacks against machine learning (ML) \cite{tuptuk2018security}. Among these, adversarial attacks against ML became extremely important since ML methods have great success in I-IoT analytics, and have become more widespread \cite{gungor2022res}. We present a workflow of how I-IoT adversarial attacks take place in Figure \ref{attack-workflow} \cite{li2020adversarial}. The workflow consists of 4 main components: (i) \textbf{System input} is the data collected from sensors in a factory environment. Different sensors can collect input data such as temperature, vibration, pressure, and proximity. (ii) \textbf{Attack} begins with the attacker exploring any vulnerability to compromise one common system input data as victim. The attacker then induces a model to craft perturbations. After tampering with the data inputted into the system back-end, the attacker can add malicious perturbations into the input data and complete the attack. (iii) \textbf{Defense} refers to protection and mitigation strategies against attacks. There are a variety of defense methods in the literature, e.g., data distortion, regularization, adversarial data detection, and adversarial training. Defense plays a key role in minimizing the impact of adversarial attacks. (iv) \textbf{System back-end} applies different ML methods to process the collected input sensor data to generate the desired information to control the actuators. Its role is crucial in extracting meaningful information from the collected data.

DL became extremely popular for data-driven predictive maintenance (PDM) applications due to its superior prediction performance \cite{gungor2021opelrul}. However, an adversary can create carefully perturbed examples to impact DL prediction performance. To effectively generate those instances, attacker can use 3 different methods \cite{warr2019strengthening}: (i) \textit{white box} methods exploit complete knowledge of a DL model, i.e., model parameters and architecture, (ii) \textit{limited black box} methods refine adversarial input based on an output generated from the model, and (iii) \textit{score-based black box} methods refine adversarial input based on the raw predictions (class probabilities) returned from the model. In this work, we use 3 different white box attack methods: fast gradient sign method (FGSM) \cite{goodfellow2014explaining}, basic iterative method (BIM) \cite{kurakin2016adversarial}, and momentum iterative method (MIM) \cite{dong2018boosting}:
\begin{itemize}
    \item \textbf{Fast Gradient Sign Method (FGSM):} FGSM first calculates the gradient of the cost function with respect to the input of the neural network. Adversarial examples are then created based on the gradient direction: $\ddot{x} = x + \epsilon * sign(\nabla_{x} \mathcal{L}(\theta, x, y))$ where $\ddot{x}$ represents the crafted adversarial examples and $\epsilon$ denotes the amount of the perturbation. 
    \item \textbf{Basic Iterative Method (BIM):} BIM applies FGSM multiple times with really small step size \cite{kurakin2016adversarial}. BIM perturbs the original data in the direction of the gradient multiplied by the step size $\alpha$: $\ddot{x} = x + \alpha * sign(\nabla_{x} J(\theta, x, y))$ where $\alpha$ is obtained by dividing the amount of perturbation ($\epsilon$) by the number of iterations ($I$): $\alpha = \epsilon / {I}$. Then, BIM clips the obtained time series elements to ensure that they are in the $\epsilon$-neighborhood of the original time series.  
    \item \textbf{Momentum Iterative Method (MIM):} MIM solves underfitting and overfitting problems in FGSM and BIM by integrating momentum into the BIM \cite{dong2018boosting}. At each iteration $i$, the variable $g_{i}$ gathers the gradients with a decay factor $\mu$: $ g_{i+1} = \mu * g_{i} + \tfrac{\nabla_{x} J(\theta, \ddot{x}_{i}, y)}{\|\nabla_{x} J(\theta, \ddot{x}_{i}, y)\|_{1}}$
    where the gradient is normalized by its $L_{1}$ distance. The perturbed data for the next iteration is created in the direction of the sign of $g_{i+1}$ with a step size $\alpha$: $\ddot{x}_{i+1} = \ddot{x}_{i} + \alpha * sign(g_{i+1})$
\end{itemize}

Other than attack generation methods, an adversary also needs to determine how to conduct an adversarial attack for a real-world system based on access level to the target model. We can categorize real-world attack patterns under 3 classes \cite{warr2019strengthening}: (i) \textit{direct attacks} allow an adversary to submit inputs to the actual target and receive corresponding results, (ii) \textit{replica attacks} use an exact replica of the target model to refine the adversarial input, (iii) \textit{transfer attacks} select a substitute model which is a good-enough approximation of the target and use this model to craft adversarial examples. Among these three, \textit{transfer attack} is the most realistic attack strategy where an adversary may not have access to the target model. \textit{Transferability property} makes these attacks dangerous where adversarial examples created on a substitute model can be effective on the target model even though these models are completely different. We take advantage of transfer attacks to propose a modified adversarial retraining approach. 

In order to detect and defend cyber-physical systems against adversarial attacks, there are distinct defense approaches proposed in the literature, combined under 3 groups \cite{warr2019strengthening, li2020adversarial}: (i) input defense, (ii) adversarial attack detection, and (iii) model defense. \textit{Input defense} pre-processes input data to remove any adversarial component before the system back-end processes it. Data compression \cite{rosenberg2019defense}, data coding \cite{rajaratnam2018speech}, data decomposition \cite{hao2015sparse}, and adding noise \cite{yuan2018commandersong} are some example input defense strategies. \textit{Adversarial attack detection} aims to distinguish attacked data from normal ones before model training and inference. Although it has been shown that these examples are not easy to detect \cite{carlini2017adversarial}, there are two different methods for adversarial attack detection: (i) \textit{data feature analysis} utilizes data statistics analysis methods such as generalized likelihood ratio test \cite{kosut2010malicious}, maximum mean discrepancy \cite{grosse2017statistical}, and temporal consistency \cite{yang2018towards}, (ii) \textit{ML-based detection} first extracts features and then train a decision model to determine if the input is an adversarial or not. Different ML methods have been used so far, e.g., one-class classifiers \cite{santana2021detecting}, extreme learning machine \cite{yan2019attack}, and reinforcement learning \cite{zhou2019secure}. We can categorize these methods under two main groups \cite{yan2019attack}: supervised and semi-supervised learning. While supervised training includes both normal and perturbed data, semi-supervised ML model is trained only using legitimate input. It might be hard or infeasible to collect adversarial data for the training process, that is why semi-supervised training provides a more realistic learning. In our work, we use semi-supervised ML-based attack detection due to its great success in detecting attacks \cite{santana2021detecting}. Santana et al. \cite{santana2021detecting} use one class support vector machine and local outlier factor to detect adversarial attacks targeting a photovoltaic power plant, yet they did not propose any defense strategy. 

\textit{Model defense} is the last category of defenses which strengthen the model itself against adversarial attacks, i.e., increasing the robustness. This is the most heavily studied defense approach where there are numerous approaches such as gradient masking \cite{papernot2017practical}, defensive distillation \cite{papernot2016distillation}, generative adversarial network \cite{akhtar2018threat}, ensemble learning \cite{gungor2022stewart}, certified defenses \cite{raghunathan2018certified}, and adversarial training \cite{bai2021recent}. \textit{Adversarial training} proves to be the most effective against adversarial attacks \cite{pang2020boosting}. The idea in adversarial training is to augment training data with adversarial examples to increase the model robustness. There are different adversarial training methods proposed: adversarial regularization \cite{zhang2019theoretically}, curriculum-based adversarial training \cite{zhang2020attacks}, ensemble adversarial training \cite{gungor2022dense}, and adaptive adversarial training \cite{cheng2020cat}. Adversarial training is formulated as a min-max problem where solving the inner maximization problem is a challenge. To create perturbed examples and incorporate them into the adversarial training, different attack generation methods can be used, e.g., fast gradient sign method \cite{kurakin2018adversarial}, projected gradient descent \cite{madry2017towards}. Hence, the quality of the included examples plays a key role in model robustness. In this work, we modify adversarial training process by considering a more realistic transfer attack scenario where the attacker does not access to the training data both for substitute and target models. We train our substitute model on a different dataset and transfer the model to be used at a different dataset. Here, we create perturbed training examples and include them in adversarial retraining process of the target model to increase its robustness. Overall, this work has 4 important contributions that are missing in the state-of-the-art: (i) accurate attack detection approach, (ii) novel and realistic adversarial retraining strategy, (iii) selective training mechanism as an additional defense layer that can select adversarial retraining or standard training based on adversarial attack detection output, and (iv) effective defense mechanism to minimize the impact of adversarial attacks.       

\section{DOuble DEfense Mechanism Framework (DODEM)}
Figure \ref{framework} demonstrates our double defense mechanism framework (\textit{DODEM}). Given multi-variate time-series sensor data, we first divide it into training and test sets. In our first defense mechanism, adversarial attack detection, we provide the training data without any attack, yet include perturbed examples in the test data. For adversarial attack detection, we use two different novelty detection algorithms: one-class support vector machine (OCSVM) and local outlier factor (LOF). In training these algorithms, we provide our training data and extract features using 3 different techniques: statistical features, chaos-theoretic measures, and stacked denoising autoencoder. These features are then provided to the selected methods and we obtain the trained classifiers. For the test data, we use the same feature extraction methods and give those to the trained classifiers. As an output, we obtain the labels representing if a specific test instance is an adversarial or normal example. Based on this class output, we continue with the selective training module, our second defense mechanism, which decides if we should choose adversarial retraining or standard training. The selective training module is motivated by the fact that adversarial retraining increases the model robustness on adversarial examples, yet decreases accuracy on clean data. If an adversarial attack is detected, then adversarial retraining process starts. If there is no attack detected, then we select standard training where we simply train target models given training data. By making this selection, our goal is to minimize the impact of adversarial attacks and protect clean data accuracy at the same time.

\begin{figure}[t]
\centering
\includegraphics[width=0.8\linewidth]
{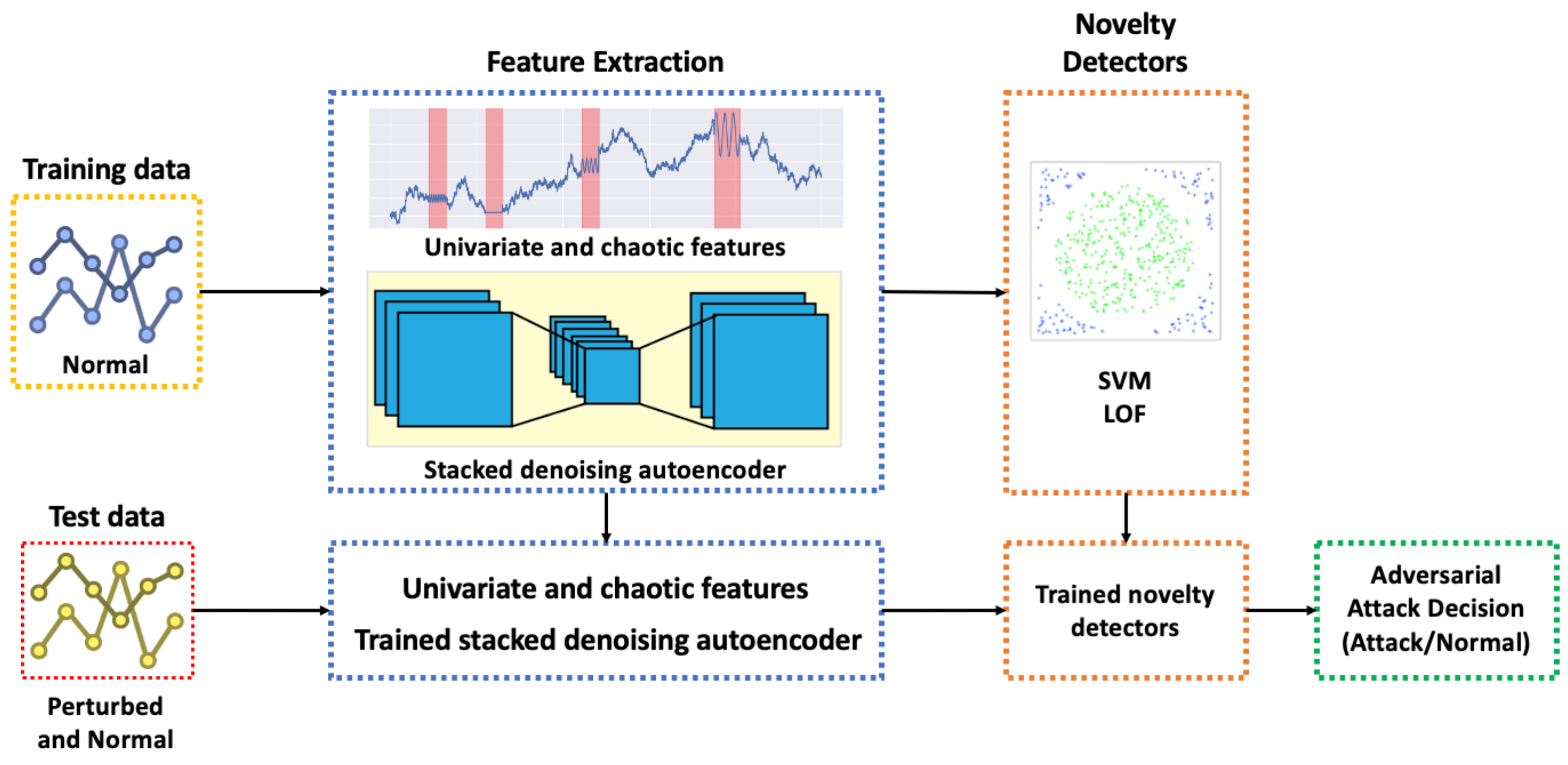}
\caption{Adversarial Attack Detection}
\label{attack-detection}
\end{figure}

\subsection{Adversarial Attack Detection}
Figure \ref{attack-detection} illustrates our adversarial attack detection module. After input sensor data is divided into training and test, we introduce adversarial examples solely for the test data. Here, we assume that the adversary picks random test instances to perform an adversarial attack. After salient features are extracted for training data, one-class classifiers are trained. Test data features are then provided to obtain if a random test instance belongs to adversarial or normal data. Different than the state-of-the-art, we use an extensive and diverse feature extraction methods and train two different novelty detection algorithms, increasing the generalizability of our study. The adversarial attack detection module has 3 main components which are feature generation, model training, and testing: 

\textbf{1) Feature generation:} Feature generation (extraction) is an important process in developing predictive analytical solutions. There are different feature generation methods \cite{yan2019attack, abdu2020detecting, santana2021detecting}: denoising autoencoder, uni-variate/multi-variate features, chaos-theoretic measures, and physical features. Our goal is to generate salient features that can better discriminate attacks from normal data. We perform our feature calculations using a sliding window to capture the temporal effects appropriately. We have $n$ sensor measurements, $\zeta^{1}, \zeta^{2}, \dots ,\zeta^{n}$ and a sliding window size of $w$. We use three categories of features, namely statistical uni-variate, chaos-theoretic measures, and denoising autoencoder.  

\textbf{a) Uni-variate features:} For each individual measurement $\zeta^{i}$, we can formulate windowed segment of measurements at time $t$ as $\psi_{t}^{i} = \zeta_{t-w}^{i}, \zeta_{t-w+1}^{i}, \dots, \zeta_{t}^{i}$. We calculate 8 features for $\psi_{t}^{i}$: minimum, maximum, mean, median, standard deviation, range, mean to maximum ratio, and minimum to maximum ratio.

\textbf{b) Chaos-theoretic measures:} Chaos theory focuses on underlying patterns and deterministic laws of dynamical systems. There are two main reasons why we utilize chaos-theoretic measures \cite{abdu2020detecting}: (i) although the perturbations are small, they contribute to the complexity of the signal, and (ii) since the perturbations are seemingly chaotic, they may be observable using measures of chaos. For the first reason, we adopt the Sample Entropy \cite{richman2000physiological} which quantifies the complexity of time-series data. For the second reason, we select 3 different features: (i) Detrended Fluctuation Analysis (DFA) \cite{peng1994mosaic} is a measure of the statistical self-affinity of a non-stationary signal, (ii) Hurst exponent \cite{hurst1957suggested} is a measure of the long-term memory of a time series, and (iii) Lyapunov exponent \cite{rosenstein1993practical} measures chaos and unpredictability. Overall, we have 4 chaos-theoretic measures.

\textbf{c) Denoising autoencoder:} In deep representation learning, one can stack up multiple layers of shallow learning blocks, train it, and use the last layers to learn useful representations of input data. Autoencoder (AE) is a type of neural network that is commonly used for representation learning. It consists of encoder and decoder sub-models where the encoder compresses the input and the decoder attempts to recreate the input from the compressed version provided by the encoder. To prevent AE to learn trivial hidden representations, denoising AE (DAE) add noise or mask some of the input values. Stacked denoising autoencoder (SDAE) \cite{vincent2010stacked} stack up individual DAEs to form a deep network where the outputs of the hidden neurons at the lower layer of DAE is the input to the upper layer of DAE. We use the hidden representations at the last layer of the SDAE to obtain 25 features.

\begin{figure}[t]
\centering
\includegraphics[width=0.45\linewidth]
{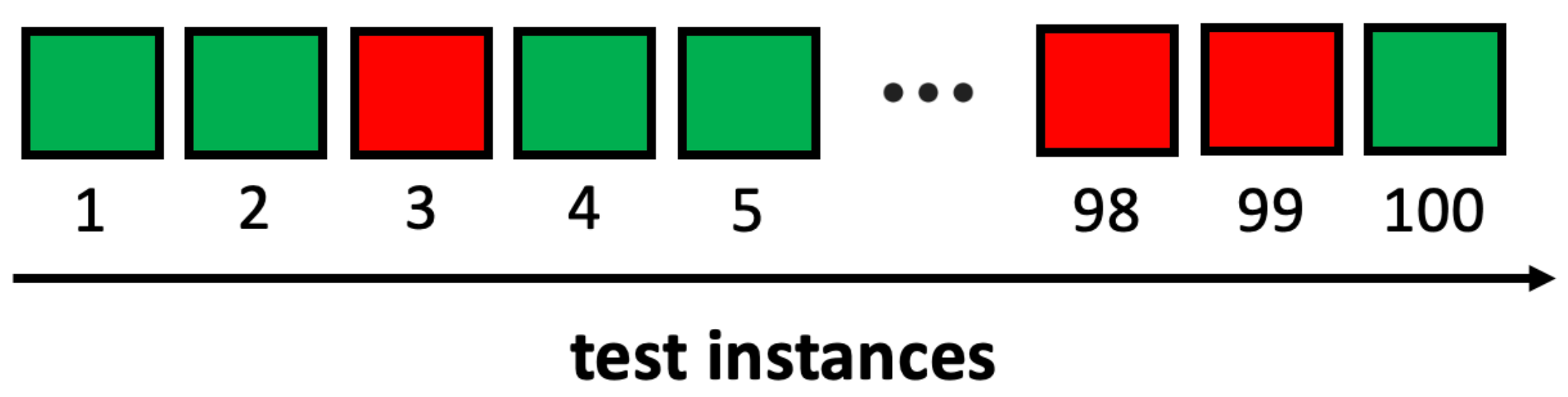}
\caption{Random Adversarial Attack Simulation}
\label{random-selection}
\end{figure}

\textbf{2) Model training:} The goal of model training is to map extracted features to binary decisions (attack or normal). We frame this problem as \textit{novelty detection} since the quantity of adversarial data is significantly smaller than normal data. The problem of novelty detection can be seen as one-class classification where one class has to be distinguished from all other possibilities \cite{moya1993one}. In our work, we utilize two different one-class classification algorithms, namely one-class support vector machine and local outlier factor:

\textbf{a) One-class Support Vector Machine (OCSVM):} OCSVM is a variation of traditional SVM where instead of using a hyperplane to separate two classes of instances, OCSVM finds the smallest hypersphere to encompass all of the instances \cite{scholkopf1999support}. In testing time, OCSVM can identify whether an instance is within this hypersphere. Based on this, it classifies the instance as adversarial or normal.  

\textbf{b) Local Outlier Factor (LOF):} LOF measures the local deviation of a given data point with respect to its neighbors \cite{breunig2000lof}. Based on this measurement, LOF identifies data points with a lower density as novelties. In order to decide how many neighbors to consider, k-Nearest Neighbors algorithm is used.

\textbf{3) Model testing:} To evaluate the performance of our attack detection module, we simulate an attacker that creates adversarial attack samples. In order to have a realistic attack scenario, we assume that the I-IoT system does not have access to these samples, hence they are not included in model training but added to the test data set. When simulating an adversarial attack scenario, we adopt a random attack fashion as illustrated in Figure \ref{random-selection} where red squares illustrate adversarial instances while green ones are regular instances. We control the amount of adversarial samples with a parameter, i.e., adversarial sample ratio. For instance, if this ratio is 5\%, then we insert 5\% adversarial samples into the original test data set. We create those examples using FGSM due to its efficiency. After we construct the test data which consists of both normal and adversarial instances, we continue with the feature generation for which we calculate uni-variate features, chaos-theoretic measures, and use the last layer of the trained SDAE. Then, we provide these features to the trained classifiers to obtain the corresponding labels (attack/normal). We ultimately calculate the required metrics to measure the performance of our attack detection module, e.g., recall, $F_{2}$ score, ROC AUC score.

\subsection{Selective Training}
The goal of our selective training module is to select adversarial retraining or normal training based on the output from adversarial attack detection module. If an instance is classified as adversarial, then we use adversarial retraining, otherwise we are safe with standard training. We propose this selection approach because of two characteristics of adversarial retraining: (i) poor prediction performance on clean data and (ii) high robustness on adversarial data. We can eliminate the weakness of adversarial retraining by selecting standard training when there is no attack.

\begin{figure}[t]
\centering
\includegraphics[width=0.7\linewidth]
{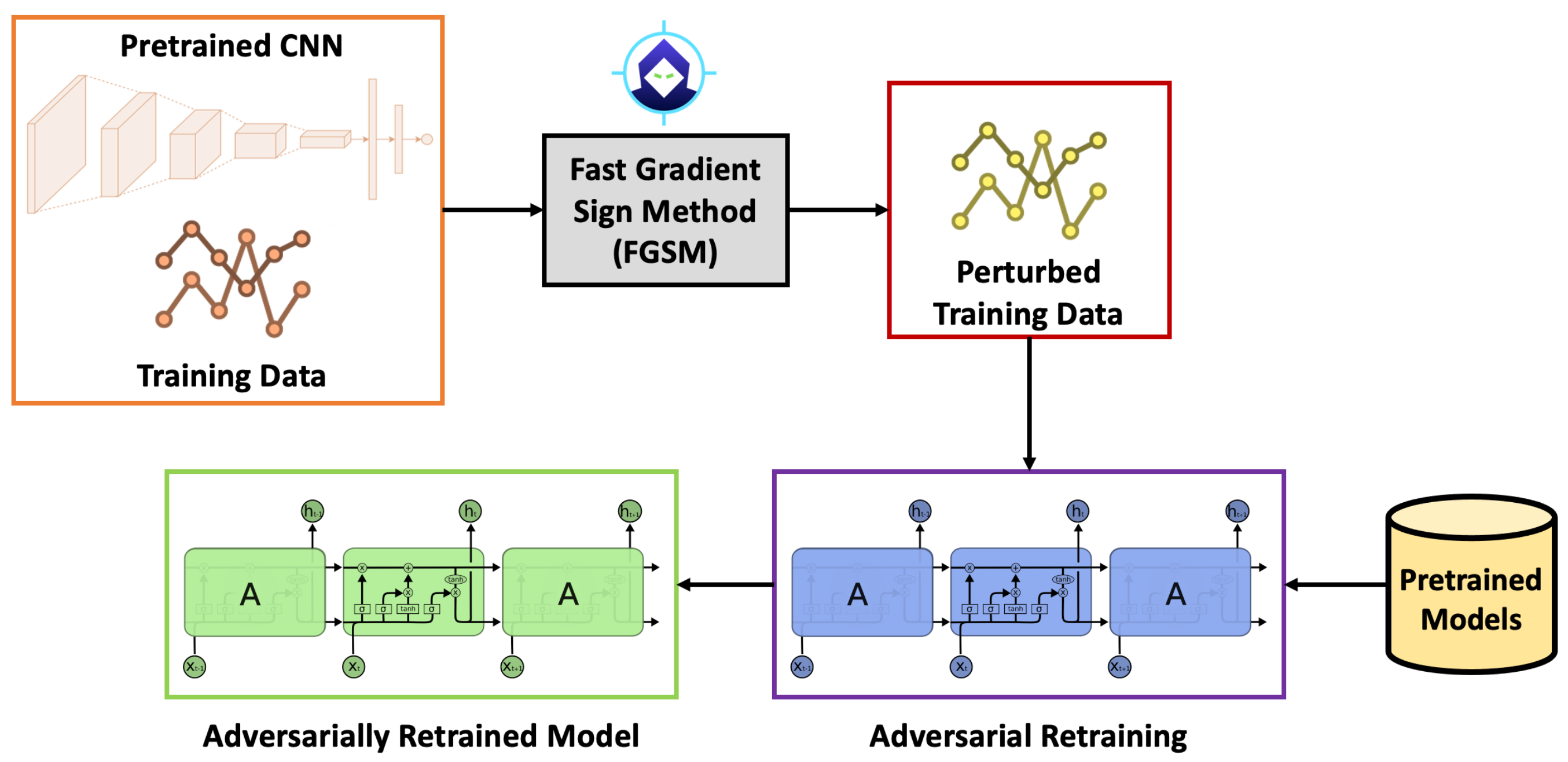}
\caption{Adversarial Retraining}
\label{adversarial-retraining}
\end{figure}

\begin{figure}[t]
\centering
\includegraphics[width=0.8\linewidth]
{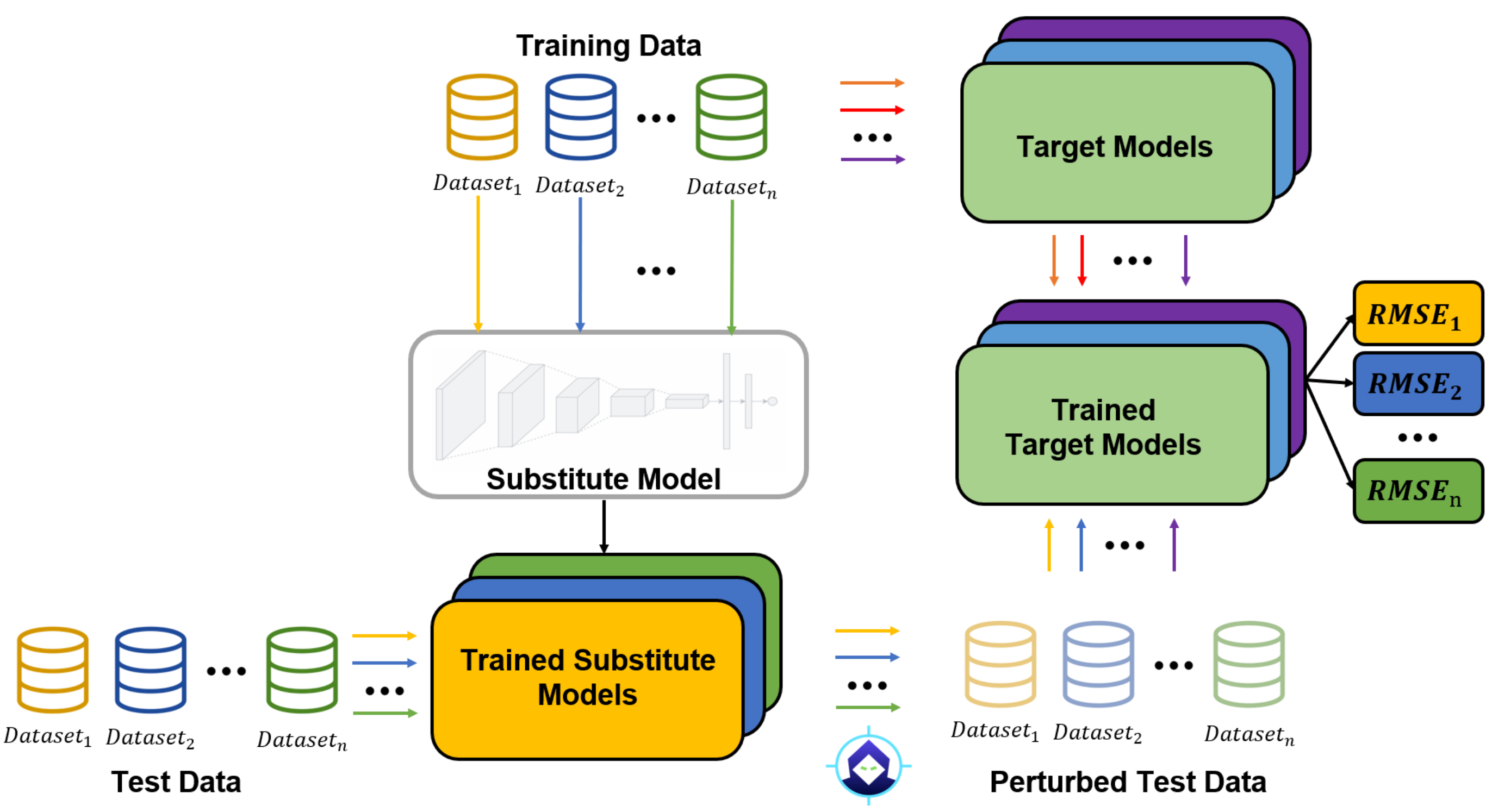}
\caption{Transferability Framework}
\label{transferability-framework}
\end{figure}

\textbf{1) Adversarial Retraining:} Figure \ref{adversarial-retraining} depicts our adversarial retraining approach. We modify the traditional adversarial training in terms of three aspects to make it more realistic: (i) we use transfer attack strategy to produce perturbed training examples to be included in adversarial retraining, (ii) in our transfer attack, we select different DL model structures as substitute and target models, (iii) we prohibit an adversary to use the same data for substitute and target model training. To modify the standard adversarial (re)training process, we first need to decide which trained model(s) to be included in adversarial retraining process. To reach that goal, we perform transferability analysis and substitute model selection: 

\textbf{Transferability Analysis:} For transfer attacks, it has been shown that two models with different architectures are likely to be susceptible to similar adversarial examples if they have been trained with the same data \cite{warr2019strengthening}. However, we might have a scenario where an adversary does not access to the same training data for substitute and target model training. In that case, we would like to test the transferability of perturbed examples. We present our transferability framework in Figure \ref{transferability-framework}. Given training data, we first train our substitute model and target models. Note that used training data is different for substitute model and target models training. Then, we create perturbed test examples using the trained substitute model via an adversarial attack since adversary does not have access to the target models. We then transfer those examples to the target models to check how much each substitute model can fool target models. We measure this based on the average prediction error value, i.e., root mean square error (RMSE), over all target models. The higher RMSE value represents that a substitute model is able to fool the target models more. 

\textbf{Substitute Model Selection:} Based on the results of the transferability analysis, we transfer the trained substitute models in an iterative manner where we start with the model that can fool the target models the most, continue with the second, and so on. To do that, we first sort the transferability analysis output error values in a descending order and start with the substitute model which gave the highest RMSE. We then craft perturbed training instances using the selected substitute model to be included in adversarial retraining. We terminate adding models under two conditions: (i) until we can no longer improve the adversarial retraining model robustness, and (ii) when we end up with the same dataset which is not allowed.

\textbf{Selected DL Models:} We select 1-D CNN as our substitute model and 7 different target models from different architectures (recurrent and hybrid), increasing the generalizability of our study:

\begin{itemize}
    \item \textbf{1-D Convolutional Neural Network (CNN):} 1D convolutional layer slides kernels across a sequence, producing a 1D feature map per kernel and each kernel learns to detect a single very short sequential pattern \cite{geron2019hands}. We adopt the 1-D CNN network proposed by Li et al. \cite{li2018remaining} which contains five consecutive CNN layers, Flatten (Dropout) layer, one fully-connected layer (with 100 nodes). 
    \item \textbf{Recurrent Neural Network (RNN):} RNN is a time-aware feedforward neural network \cite{geron2019hands}. Our network contains 3 RNN layers having 64, 32, and 16 units connected to 2 fully connected feed forward neural networks (each with 8 units). 
    \item \textbf{Long Short-Term Memory (LSTM):} LSTM has special memory cells to store information for longer. Updates in this cell can happen by the activation of three distinctive gates: 1) forget gate (the memory cell is cleared completely), 2) input gate (memory cell stores the received input), and 3) output gate (next neurons obtain the stored knowledge from the memory cell) \cite{gensler2016deep}. We adapt a similar network structure where RNN layers are replaced with LSTM layers. 
    \item \textbf{Bi-directional LSTM (BLSTM):} BLSTM also considers future data by adding a backward direction to LSTM networks \cite{wang2018remaining}. The overall network structure is similar to LSTM model where LSTM layers are replaced with BLSTM layers. 
    \item \textbf{Gated Recurrent Unit (GRU):} GRU is a simplified version of LSTM \cite{cho2014learning}. Specifically, forget and input gates are controlled by a single gate controller and there is no output gate, instead a new gate controller decides which part of the information to be transferred. We use the same network structure as LSTM except we change the LSTM layers to GRU.   
    \item \textbf{Bi-directional GRU (BGRU):} Similar to BLSTM, BGRU takes future data into consideration \cite{she2021bigru}. We construct this model by simply replacing BLSTM layers with BGRU layers.
    \item \textbf{CNN-GRU (CGRU):} We connect our CNN and GRU networks in parallel. On one path, we have our 1-D CNN architecture, on another path we have our GRU model. These are then connected to fully connected NN with 100 units.
    \item \textbf{GRU-LSTM (GLSTM):} We combine our GRU and LSTM models in parallel. GRU and LSTM paths are concatenated and connected to fully connected NN with 100 units.
\end{itemize}

For the sake of simplicity, we show substitute and target model training in Figure \ref{transferability-framework}. In Figure \ref{adversarial-retraining}, pretrained CNN corresponds to the substitute model that was able to fool the target models the most. In Section \ref{adversarial-retraining-exp}, we provide an experimental analysis how we select this model. Selected pre-trained CNN and training data, adversary first creates perturbed training instances using FGSM. 
These created examples are then used in retraining of the target models. As an output, we obtained the adversarially retrained target models. 

\textbf{2) Standard Training:} We select standard training if no attack is detected for a specific instance. This is a pretty standard training process where training data is used to train different target models.   

\section{Experimental Analysis}
\begin{figure}
\centering\includegraphics[width=0.7\linewidth]
{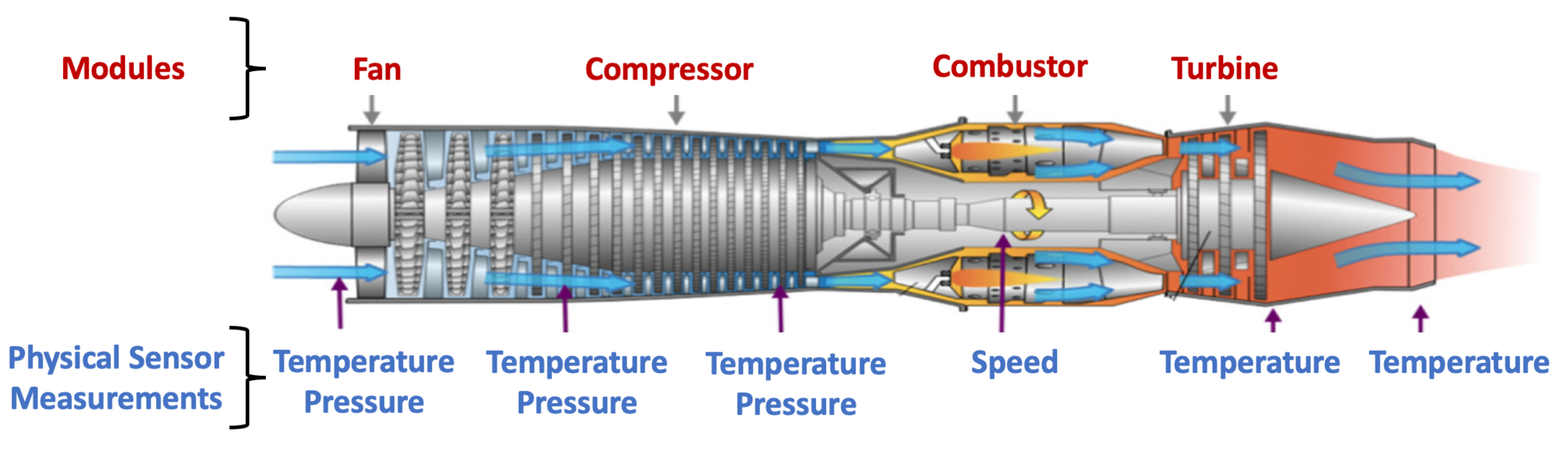}
\caption{Engine Diagram Simulated in C-MAPSS \cite{saxena2008damage}}
\label{engine-diagram}
\end{figure}

\begin{table}[t]
\centering
\caption{C-MAPSS Data Set}
\scalebox{0.9}{
\begin{tabular}{|c|c|c|c|c|}
\hline
\textbf{Data Set}                 & \textbf{FD001} & \textbf{FD002} & \textbf{FD003} & \textbf{FD004} \\ \hline
\textbf{Train trajectories}       & 100            & 260            & 100            & 249            \\ \hline
\textbf{Test trajectories}        & 100            & 259            & 100            & 248            \\ \hline
\textbf{Max/Min cycles for train} & 362/128        & 378/128        & 525/145        & 543/128        \\ \hline
\textbf{Max/Min cycles for test}  & 303/31         & 367/21         & 475/38         & 486/19         \\ \hline
\textbf{Operating conditions}     & 1              & 6              & 1              & 6              \\ \hline
\textbf{Fault conditions}         & 1              & 1              & 2              & 2              \\ \hline
\end{tabular}}
\label{data-set-cmapss}
\end{table}

\subsection{Dataset Description}
NASA Commercial Modular Aero-Propulsion System Simulation (C-MAPSS) \cite{saxena2008damage} is a benchmark dataset for remaining useful life (RUL) prediction. This dataset consists of multiple aircraft engines simulated under various operating and fault conditions. Figure \ref{engine-diagram} depicts the simplified version of simulated engine diagram and its four major components: fan, turbine, compressor, and combustor. The data is collected using different sensors (e.g. temperature, pressure) placed on these components. NASA C-MAPSS includes 4 different datasets in increasing complexity: FD001$\sim$FD004. Table \ref{data-set-cmapss} presents the dataset and their corresponding features such as number of train and test trajectories, maximum/minimum number of cycles, etc. We observe that FD001 is the simplest data set and FD004 is the most complicated one with the highest number of operating and fault conditions. For each dataset, we have separate training and test data. The training data contains the entire lifetime of an engine and test data is terminated at some point before engine failure. The goal is to predict RUL for the test data. C-MAPSS feature columns include the engine ID, cycle index, three operational settings, and 21 sensor measurements.

\subsection{Experimental Setup}
We run all experiments on a PC with 16 GB RAM and an 8-core 2.3 GHz Intel Core i9 processor. For adversarial attack detection, and adversarial retraining, we use FGSM since it can create adversarial examples efficiently, i.e., fast and less detectable adversarial example generation \cite{wong2020fast}. In order to test \textit{DODEM} against adversarial attacks, we use FGSM, basic iterative method (BIM), and momentum iterative method (MIM) \cite{fawaz2019adversarial, dong2018boosting}: amount of perturbation($\epsilon$)=0.1, step size($\alpha$)=0.001, number of iterations($I$)=100, decay factor($\mu$)=1.    

\textbf{Feature Generation:} For statistical uni-variate features, we use the NumPy library. For the chaos-theoretic measures, we use the nolds module in Python. Lastly, we use a 2-layer SDAE network, and the network structures for the first-layer DAE and second-layer DAE are 20-50-20 and 50-25-50, respectively, where 50 hidden neuron outputs of the first-layer DAE are used as the inputs to the second-layer DAE. The outputs of the 25 hidden neurons of the second-layer DAE are taken as the features, which give us 25 learned features. 

\begin{figure}
\centering\includegraphics[width=0.6\linewidth]
{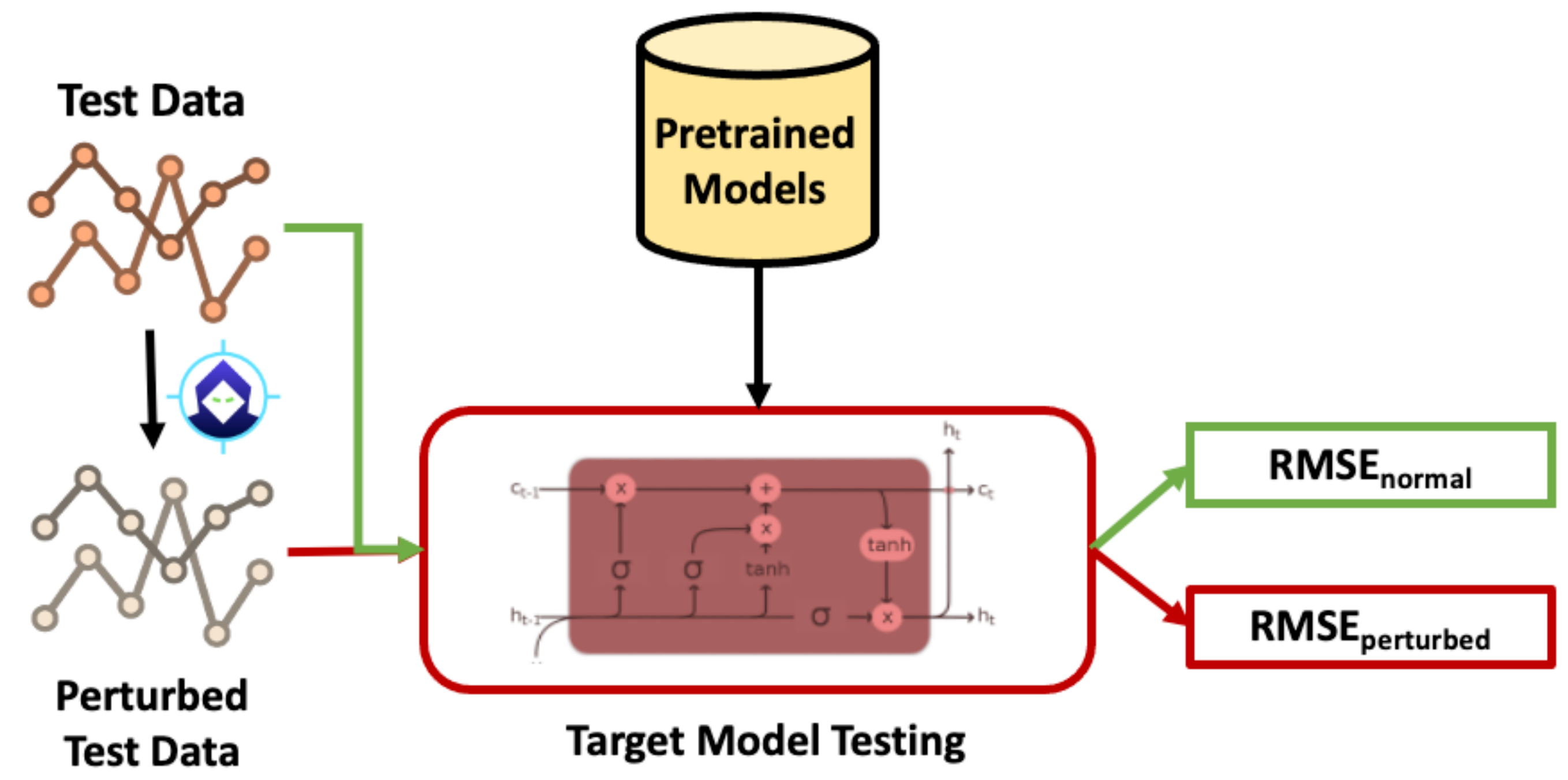}
\caption{Testing Framework}
\label{testing-framework}
\end{figure}

\textbf{Adversarial Attack Detection:} We use the scikit-learn library \cite{pedregosa2011scikit} to implement one-class classifiers. For OCSVM, we select the \textit{rbf} kernel and try different $\nu$ values from the set \{0.001, 0.005, 0.01, 0.05, 0.1, 0.2\}. For the LOF, we set the number of neighbors to 20 while selecting contamination value from the same set we used for OCSVM $\nu$ values. To measure the prediction performance of these algorithms, we use the $F_{\beta}$ score metric where we set the $\beta$ to 2 to weigh recall higher than precision: 
\begin{equation}
    F_{\beta} = (1+\beta^{2}) \cdot \dfrac{precision \cdot recall}{(\beta^{2} \cdot precision)+recall}
\end{equation}

$F_{2}$ score is a more suitable metric for attack detection because it is more important to classify adversarial samples correctly as much as possible, i.e., true positives are more important. The higher $F_{2}$ corresponds to a higher attack detection performance. We try different adversarial test ratios (\{0.01, 0.05, 0.1, 0.2\}) in our experiments. This ratio represents the amount of adversarial data within the entire test dataset. 

\textbf{DL Models:} The following hyper-parameters are selected for DL training: Adam optimizer with learning rate 0.001, elu activation function, batch size of 128, and a max number of epochs of 150 where callback is activated (patience is set to 10 for validation data), and sliding time window size of 80. For adversarial retraining, we retrain the target models for 30 more epochs. 

\textbf{Robustness Metric:} RUL prediction is a typical regression problem where input data is mapped to a continuous output value. To measure the robustness of our proposed approach against adversarial attacks, we first measure the root mean square error (RMSE) of the RUL prediction process as follows:
\begin{equation}
RMSE = \sqrt{ \dfrac{1}{\mathcal{N}} \sum_{i=1}^{\mathcal{N}} \epsilon_{i}^2} 
\end{equation}
where $\mathcal{N}$ is the number of samples, $\epsilon$ is the difference between the estimated RUL ($RUL_{est}$) and the true RUL ($RUL_{true}$). Smaller RMSE means that we have a more robust model since the model is affected less by adversarial attacks. 

\textbf{Testing Framework:} We measure the robustness of our proposed method with three white-box adversarial attack scenarios, namely FGSM, BIM, and MIM. These are 3 most common attacks for time series data \cite{abdu2020detecting}. Figure \ref{testing-framework} illustrates our white-box testing framework. Given clean test data, the adversary first crafts adversarial test samples using pre-trained target models. Clean and perturbed test data are given to the target models and we measure normal and perturbed RMSE, respectively over all target models.

\begin{figure}
\centering
\begin{subfigure}{0.4\linewidth}
    \centering
    \includegraphics[width=\linewidth]{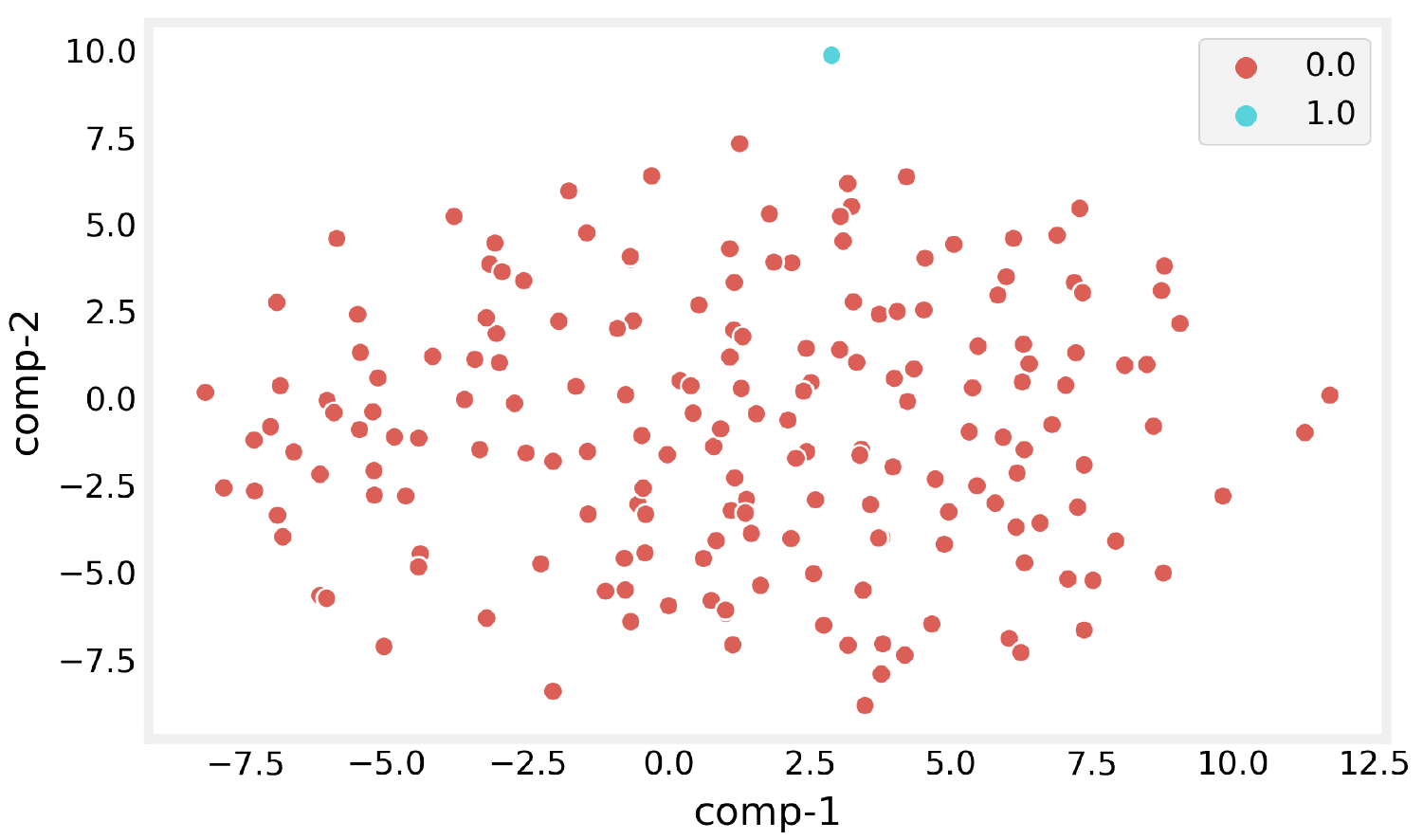}
    \caption{\textbf{Adversarial Data Ratio: 1\%}}
\end{subfigure}
\begin{subfigure}{0.4\linewidth}
    \centering
    \includegraphics[width=\linewidth]{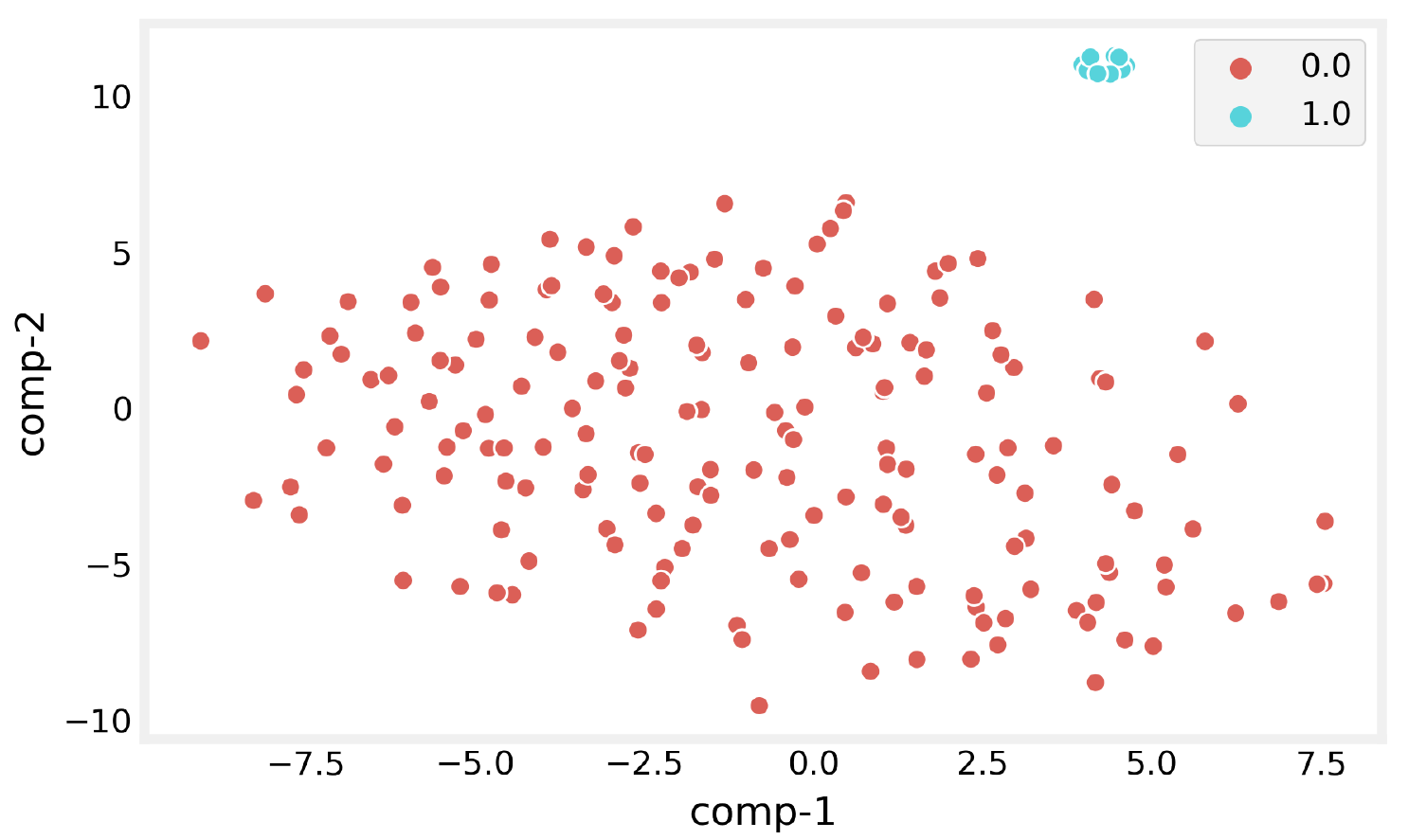}
    \caption{\textbf{Adversarial Data Ratio: 5\%}}
\end{subfigure}
\newline
\begin{subfigure}{0.4\linewidth}
    \centering
    \includegraphics[width=\linewidth]{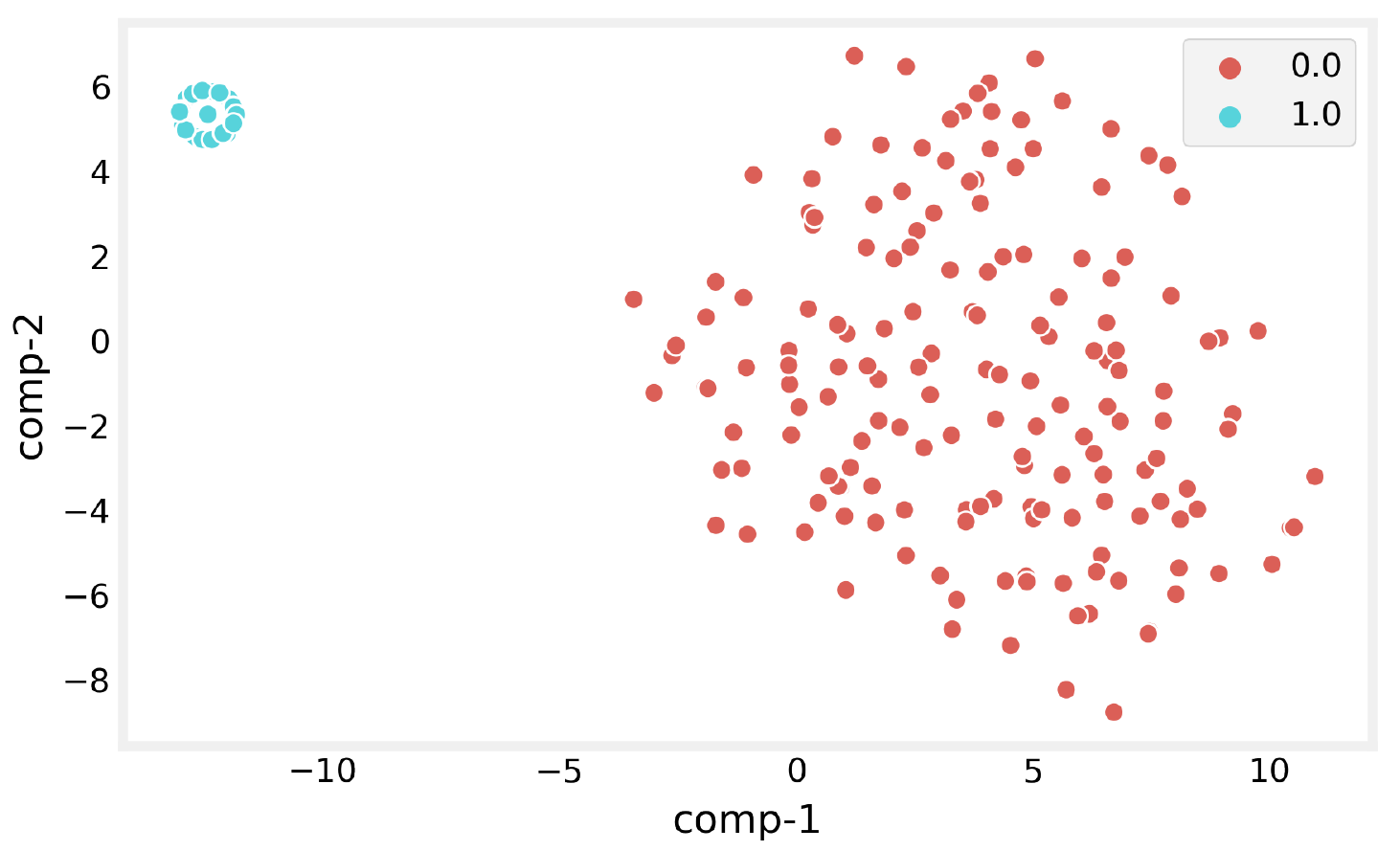}
    \caption{\textbf{Adversarial Data Ratio: 10\%}}
\end{subfigure}
\begin{subfigure}{0.4\linewidth}
    \centering
    \includegraphics[width=\linewidth]{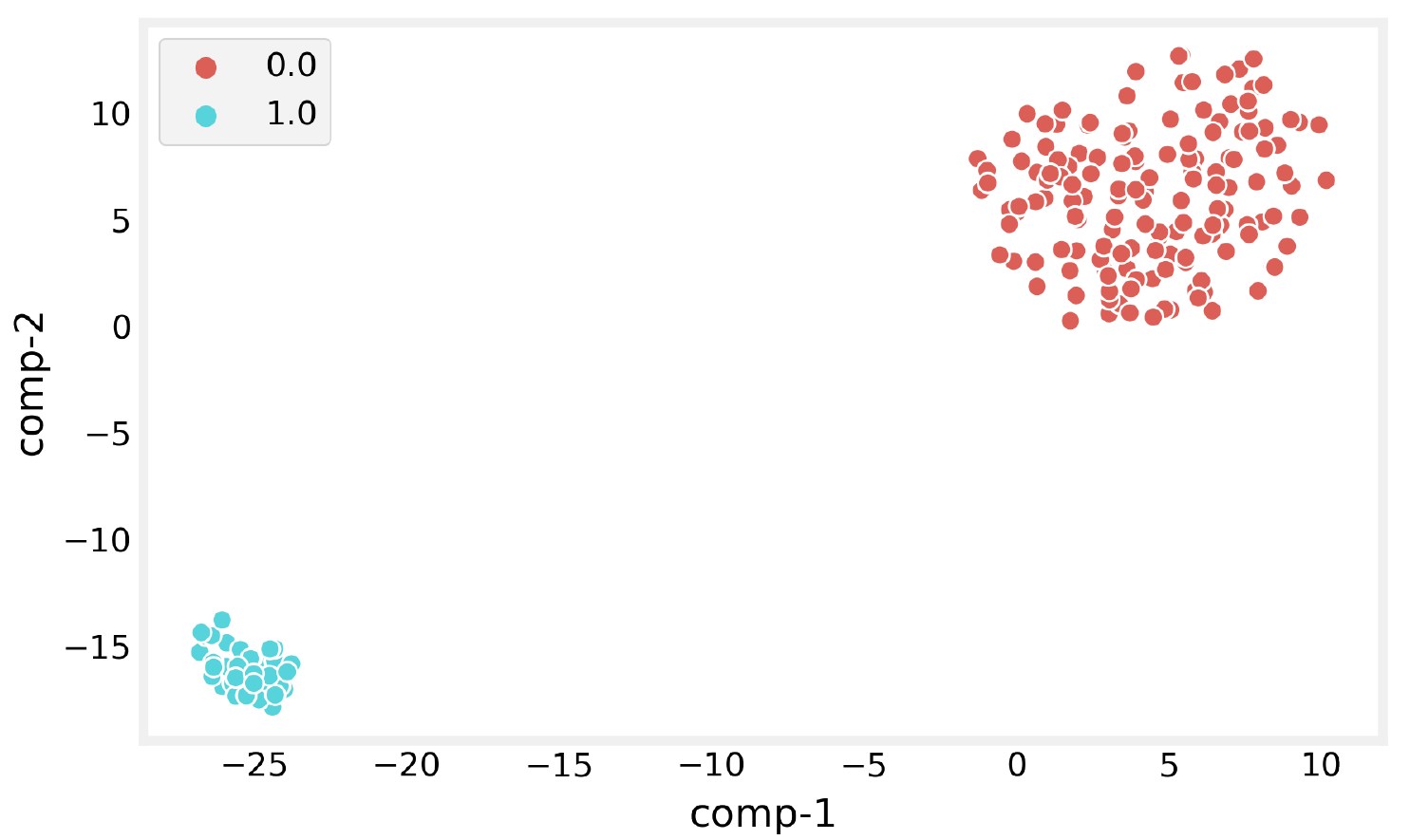}
    \caption{\textbf{Adversarial Data Ratio: 20\%}}
\end{subfigure}
\caption{FD002 t-SNE Projection}
\label{tsne-results}
\end{figure}

\subsection{Adversarial Attack Detection}
To understand how selected features are helpful in distinguishing adversarial samples from legitimate ones, we project the selected 37-D feature space to a 2-D space using t-distributed stochastic neighbor embedding (t-SNE) \cite{van2008visualizing}. Figure \ref{tsne-results} shows the t-SNE projection for FD002 (Note that we only present FD002 for clarity in this figure. We obtain similar figures for the other datasets). Each sub-figure corresponds to a different adversarial data ratio, varying from 1\% to 20\%. While red dots indicate normal data, light blue are the adversarial samples. We can see that the normal and attack samples are locally separable in the feature space. This indicates that our attack detection methods should perform well in separating normal data and and crafted adversarial data. 

\begin{figure}
\centering
\begin{subfigure}{0.4\linewidth}
    \centering
    \includegraphics[width=\linewidth]{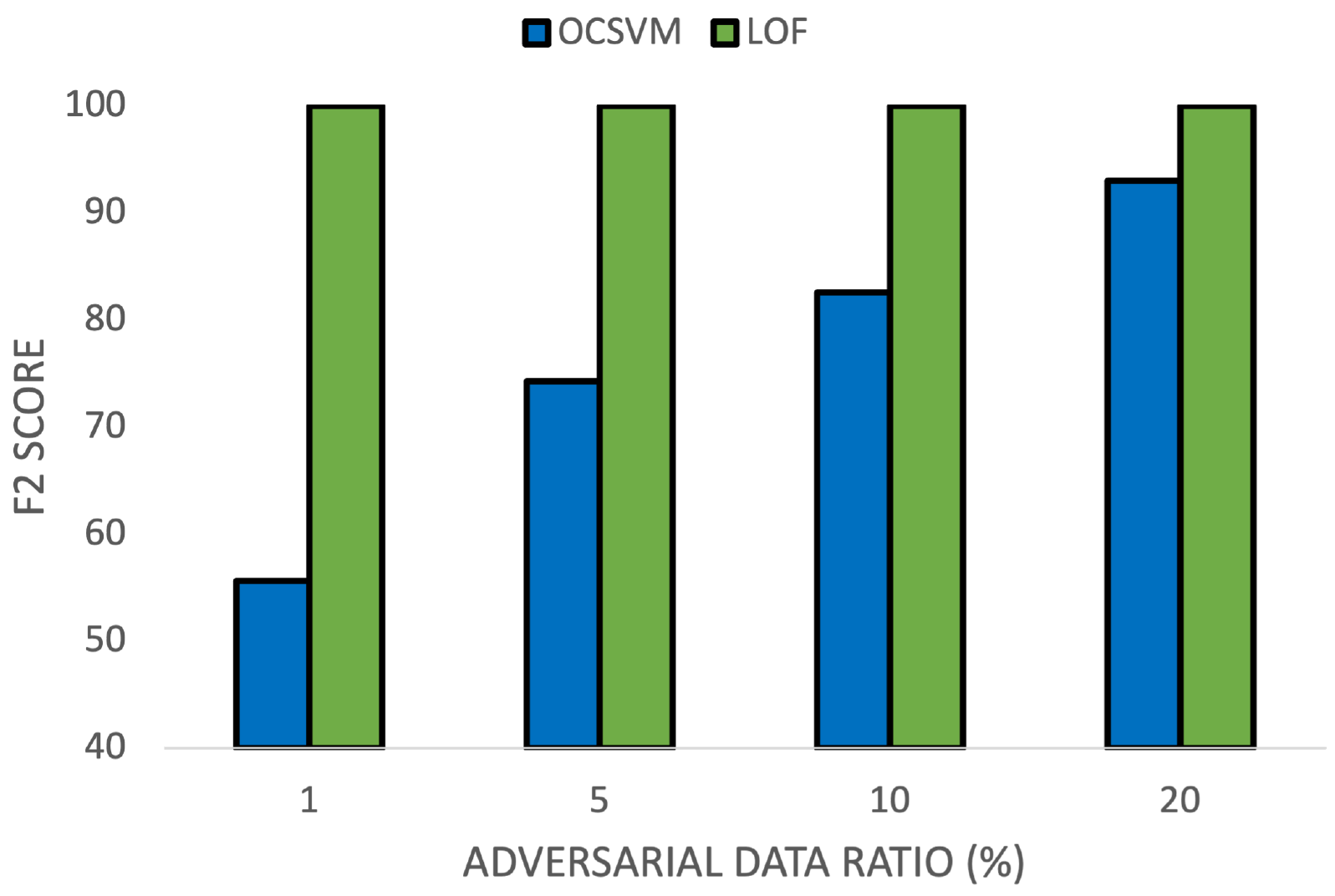}
    \caption{\textbf{FD001}}
\end{subfigure}
\begin{subfigure}{0.4\linewidth}
    \centering
    \includegraphics[width=\linewidth]{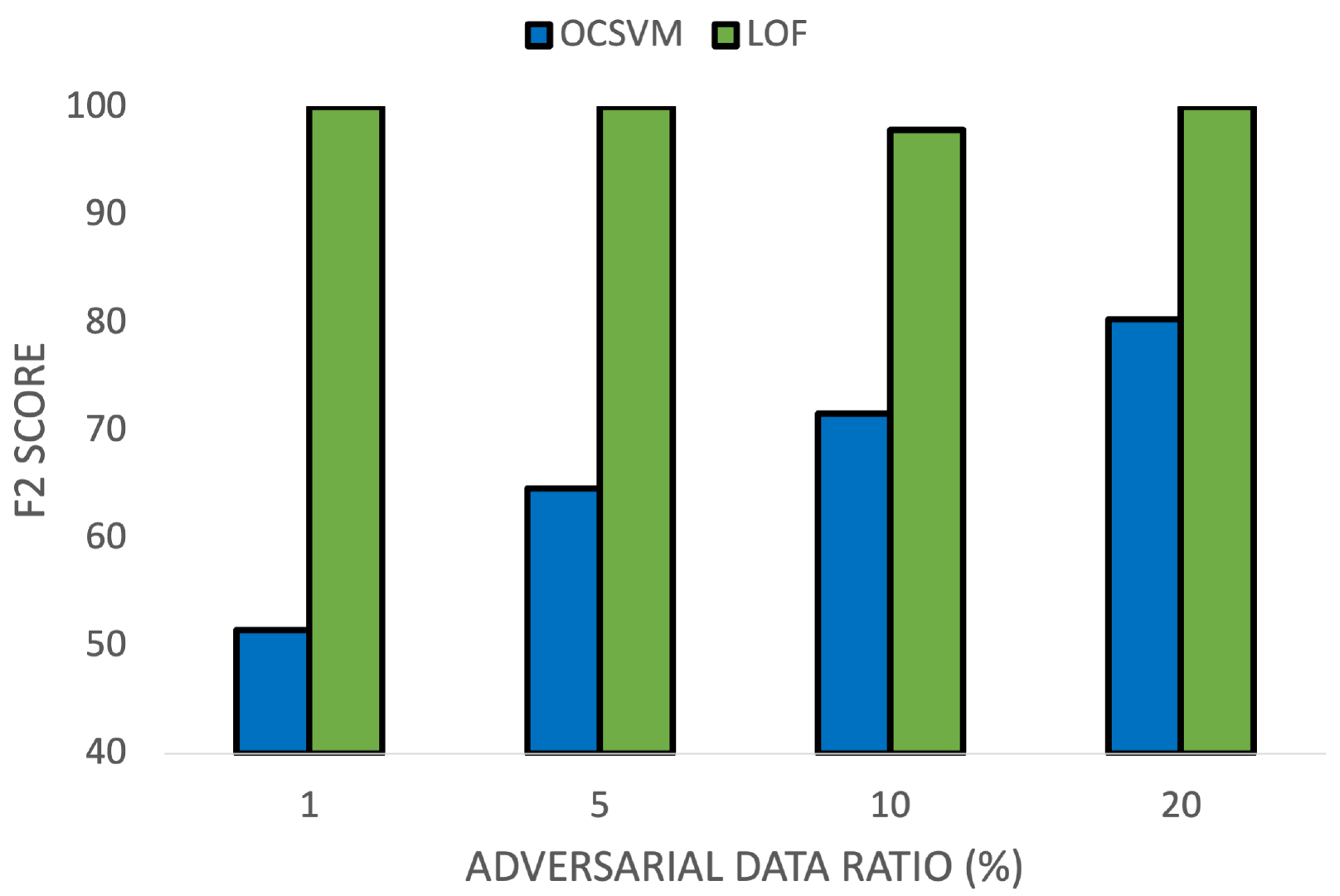}
    \caption{\textbf{FD002}}
\end{subfigure}
\newline
\begin{subfigure}{0.4\linewidth}
    \centering
    \includegraphics[width=\linewidth]{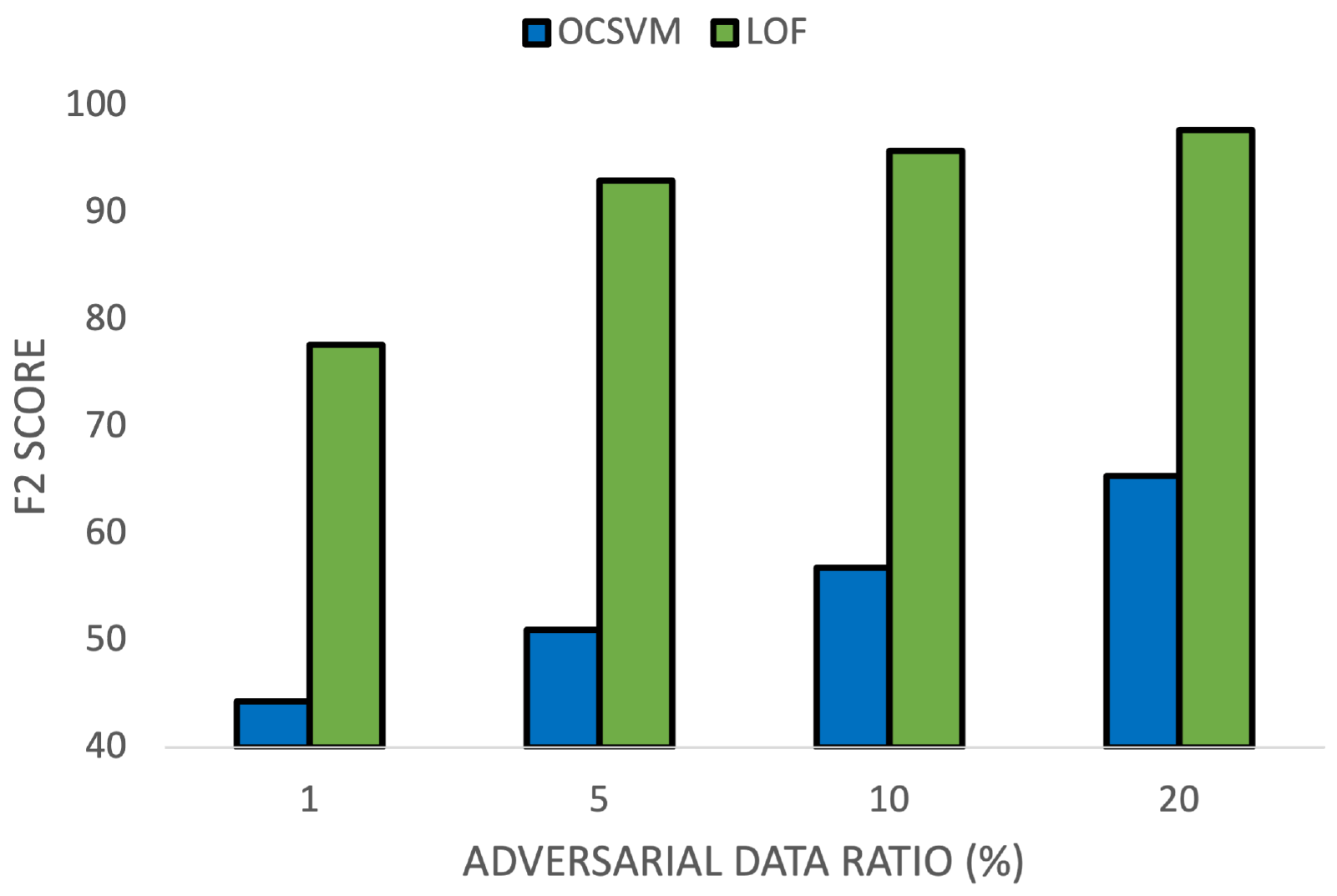}
    \caption{\textbf{FD003}}
\end{subfigure}
\begin{subfigure}{0.4\linewidth}
    \centering
    \includegraphics[width=\linewidth]{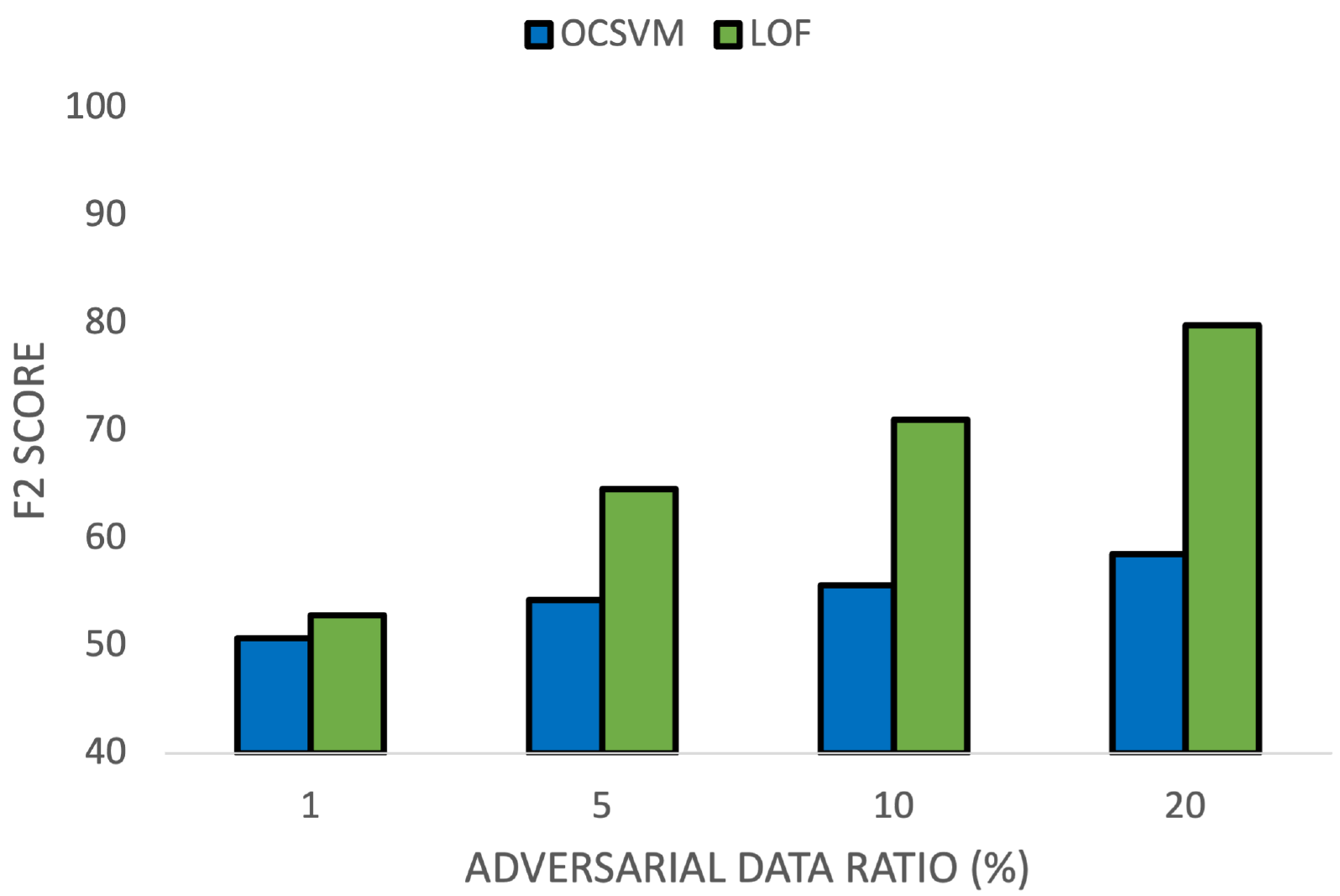}
    \caption{\textbf{FD004}}
\end{subfigure}
\caption{Adversarial Attack Detection Results}
\label{attack-detection-results}
\end{figure}

For the selected detection methods, we report the best $F_{2}$ scores among all possible $\nu$ and contamination values for OCSVM and LOF, respectively. We similarly vary the adversarial data ratio from 1\% to 20\%. Figure \ref{attack-detection-results} shows the attack detection results where each sub-figure corresponds to a different dataset. In each sub-figure, x-axis denotes the adversarial data ratio and y-axis presents $F_{2}$ scores. While blue color represents OCSVM, green corresponds to LOF. We observe that LOF significantly outperforms OCSVM at all datasets. While LOF has 89\% average $F_{2}$ score over all datasets and adversarial data ratios, OCSVM reaches up to 63\%. As adversarial data ratio gets bigger, the predictor performance becomes better. For instance at FD003, LOF has an $F_{2}$ score of 78\% and 98\% for 1\% and 20\% adversarial data ratios, respectively. The classifiers have difficulty in recognizing adversarial samples when they are extremely scarce. As the test data has more adversarial samples, distinguishing these from the normal ones becomes easier. We also observe that as the complexity of the data increases (related to number of operating and fault conditions), the attack detection performance gets worse. LOF can reach 100\% $F_{2}$ score at FD001 (simplest dataset), but only 80\% $F_{2}$ score at FD004 (most complex). Based on these observations, we will only report LOF attack detection results in Section \ref{proposed-method-robustness} as LOF has much better adversarial attack detection performance.

\subsection{Adversarial Retraining}
\label{adversarial-retraining-exp}
\begin{figure}
\centering\includegraphics[width=0.7\linewidth]
{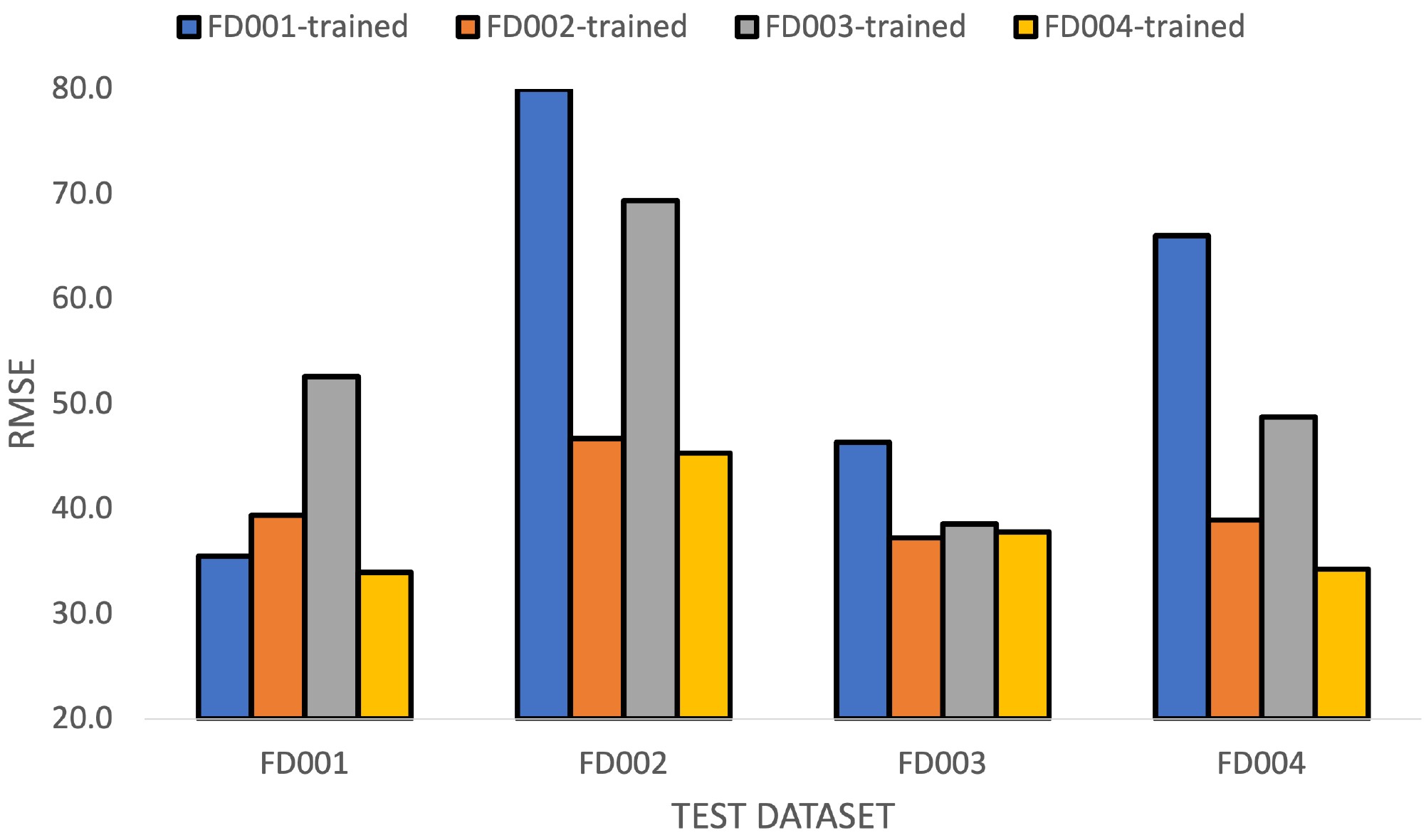}
\caption{Transferability Experiments}
\label{transfer-exps}
\end{figure}

\textbf{Transferability Experiments:} To determine which trained model(s) to include in the adversarial retraining process, we first conduct transferability experiments. The goal is to test the hypothesis that two models with different architectures are likely to be susceptible to similar adversarial examples if they have been trained with the same data \cite{warr2019strengthening}. In these experiments, we train a CNN model on four different datasets and transfer the trained model to other datasets to create perturbed test examples. We then measure the target model prediction performance under adversarial data. In total, we have 16 different scenarios, e.g., train CNN at FD001, transfer this model to FD002, FD003, and FD004. Figure \ref{transfer-exps} summarizes the  transferability results. In this figure, x-axis denotes the test data while y-axis reports the average RMSE value over all target models. Different colors correspond to CNN models trained on different datasets, for example blue denotes CNN model trained at FD001. 
For each dataset, we would expect that the RMSE values to be highest with the CNN model trained on the same data set. However, the RMSE values are worse with models that are trained on different datasets. For example, for FD002, FD001-trained CNN can fool the target models more, followed by FD003. FD001 trained CNN has an RMSE of 79.9 while this value decreases to 46.7 when same dataset is used. We can have a similar observation with other datasets as well. For FD004, FD001 and FD003 lead to worse prediction performance than FD004-trained CNN. Our transferability experiments conclude that models trained on different datasets can negatively impact the target models more. 

\textbf{Proposed Adversarial Retraining Results:} Motivated by our transferability results, we modify the traditional adversarial retraining by adding models trained on different datasets. We perform this addition iteratively where we start adding the trained model that can fool the target model the most, then the second, and so on. For example, for FD002, we add the FD001-trained CNN first (since it leads to the worst prediction performance) and then FD003-trained CNN. We stop adding models based on two conditions: (i) until we can no longer improve model robustness and (ii) when we end up with the same dataset, e.g., adversary cannot use FD002 trained model for FD002 adversarial retraining. After we determine the models to be added, we transfer these to the target dataset to craft adversarial training instances. When analyzing the validation data, we observed that the amount of training data to be used in adversarial retraining has an impact on model robustness. As a result, we choose the number of crafted adversarial samples to be 1\% of the training data size. Then, we provide these adversarial samples to the target models for retraining. We then measure RMSE under adversarial test examples. 

\begin{figure}
\centering
\begin{subfigure}{0.6\linewidth}
    \centering
    \includegraphics[width=\linewidth]{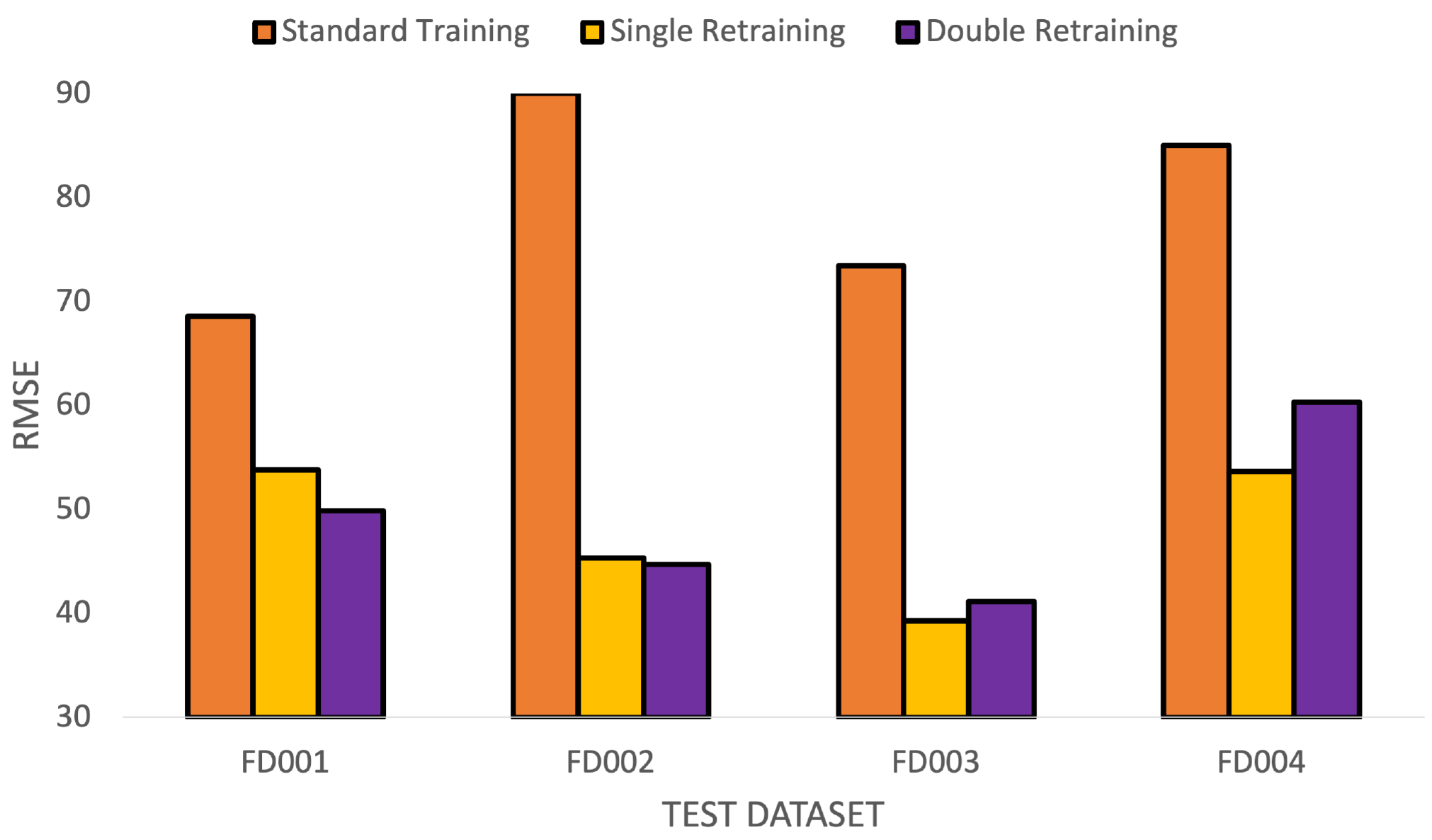}
    \caption{\textbf{Perturbed Data}}
    \label{retraining-perturbed}
\end{subfigure}
\begin{subfigure}{0.6\linewidth}
    \centering
    \includegraphics[width=\linewidth]{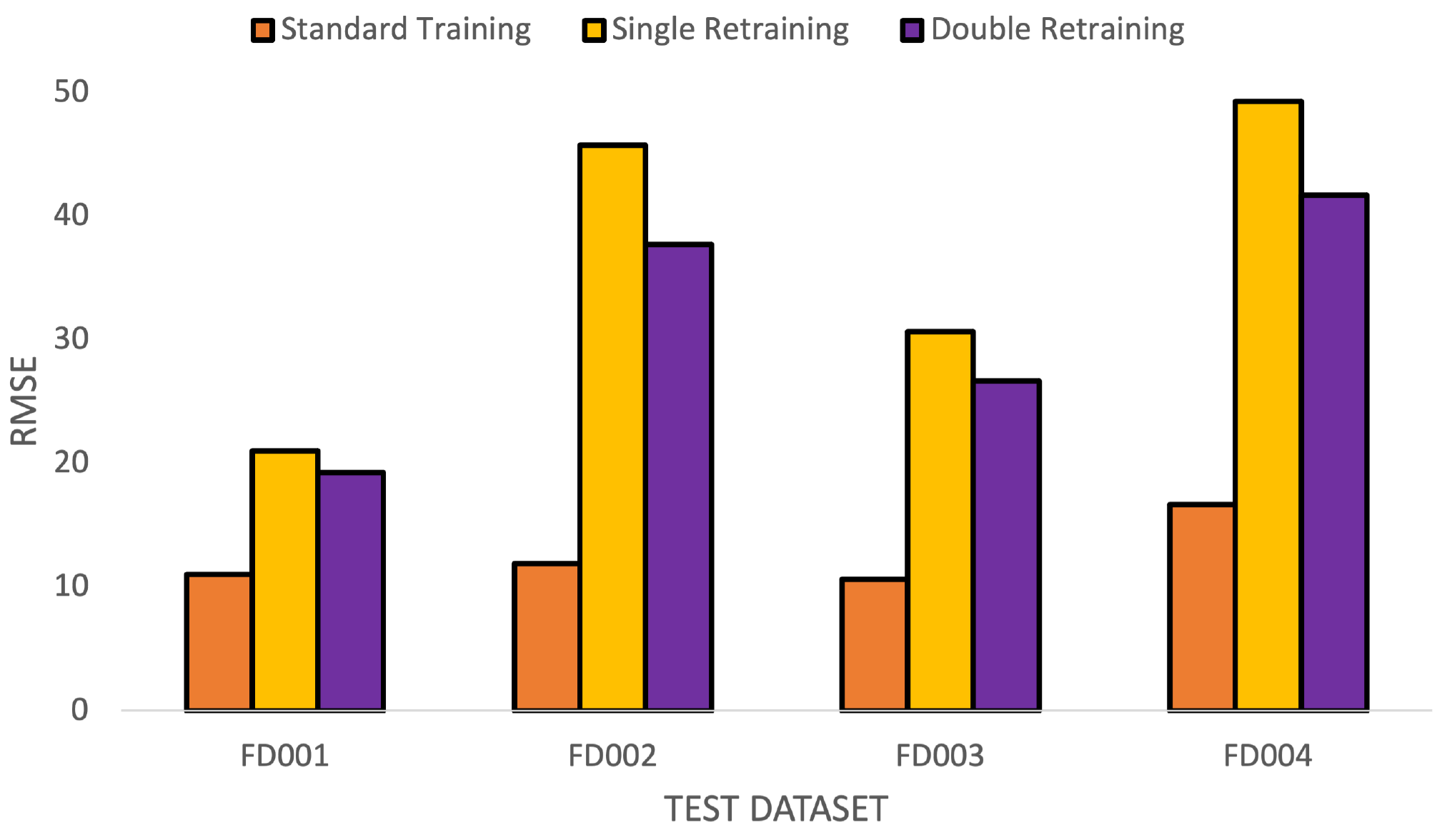}
    \caption{\textbf{Clean Data}}
    \label{retraining-clean}
\end{subfigure}
\caption{Adversarial Retraining Results}
\label{attack-retraining-results}
\end{figure}

Figure \ref{attack-retraining-results} presents the adversarial retraining results on perturbed (Figure \ref{retraining-perturbed}) and clean (Figure \ref{retraining-clean}) data compared to standard training. For both sub-figures, we have test dataset on the x-axis and average RMSE (over all target models) on the y-axis. Orange color represents standard training, yellow color denotes adversarial retraining with one model (single retraining), and purple corresponds to adversarial retraining with two models (double retraining). Here, adversarial retraining (single/double) corresponds to our proposed retraining approach. To measure the perturbed and clean data RMSE values, we consider two extreme cases. For perturbed data, we use a test dataset with only adversarial samples, for clean data our test data does not include any adversarial example. Figure \ref{retraining-perturbed} shows that our adversarial retraining approach improves the robustness significantly when there is adversarial attack. We improve model robustness by up to 67\% (40\% on average). We also observe that introducing an additional trained model into adversarial retraining, i.e., switching from single to double retraining, does not bring any significant advantage. Although double retraining brings 7\% and 1\% more robustness for FD001 and FD002 respectively, it leads to 5\% and 11\% less robustness for FD003 and FD004. When RMSE values are averaged over all datasets, single retraining in fact outperforms double retraining by 2\%. 

Figure \ref{retraining-clean} shows that adversarial retraining performance on clean data is significantly worse than standard training. Although double retraining has smaller error compared to single retraining, they both under-perform compared to standard training when using clean data. This creates an important issue for adversarial training because in a realistic scenario, we may not know which samples are targeted for an adversarial attack or if there is an attack or not at all. 
Our framework, \textit{DODEM} solves this problem by performing selective training which combines adversarial training with standard training based on adversarial attack detection output. \textit{DODEM} leads to lower error on both clean and perturbed data, providing a more robust solution.

\begin{figure}
\centering
\begin{subfigure}{0.4\linewidth}
    \centering
    \includegraphics[width=\linewidth]{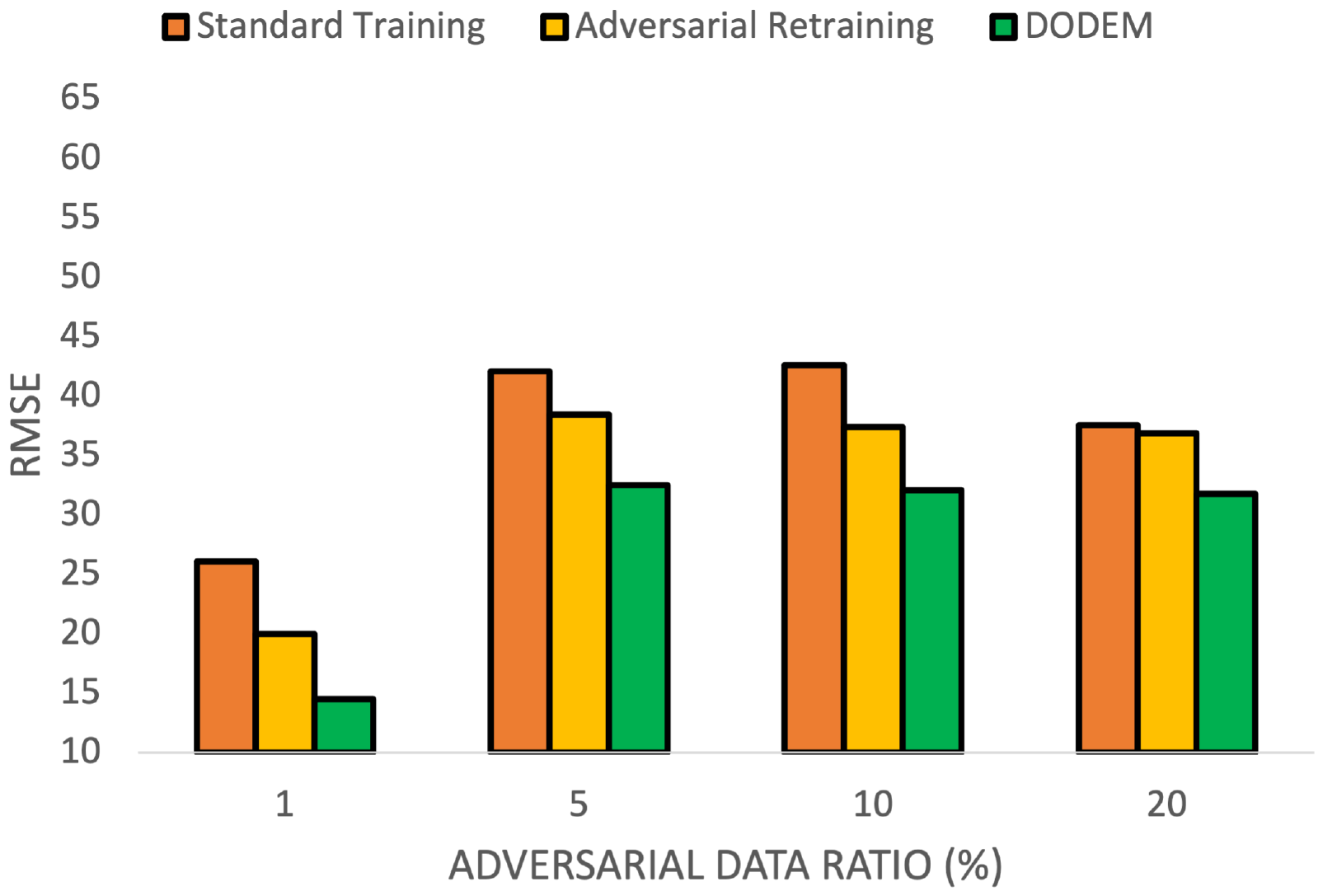}
    \caption{\textbf{FD001}}
\end{subfigure}
\begin{subfigure}{0.4\linewidth}
    \centering
    \includegraphics[width=\linewidth]{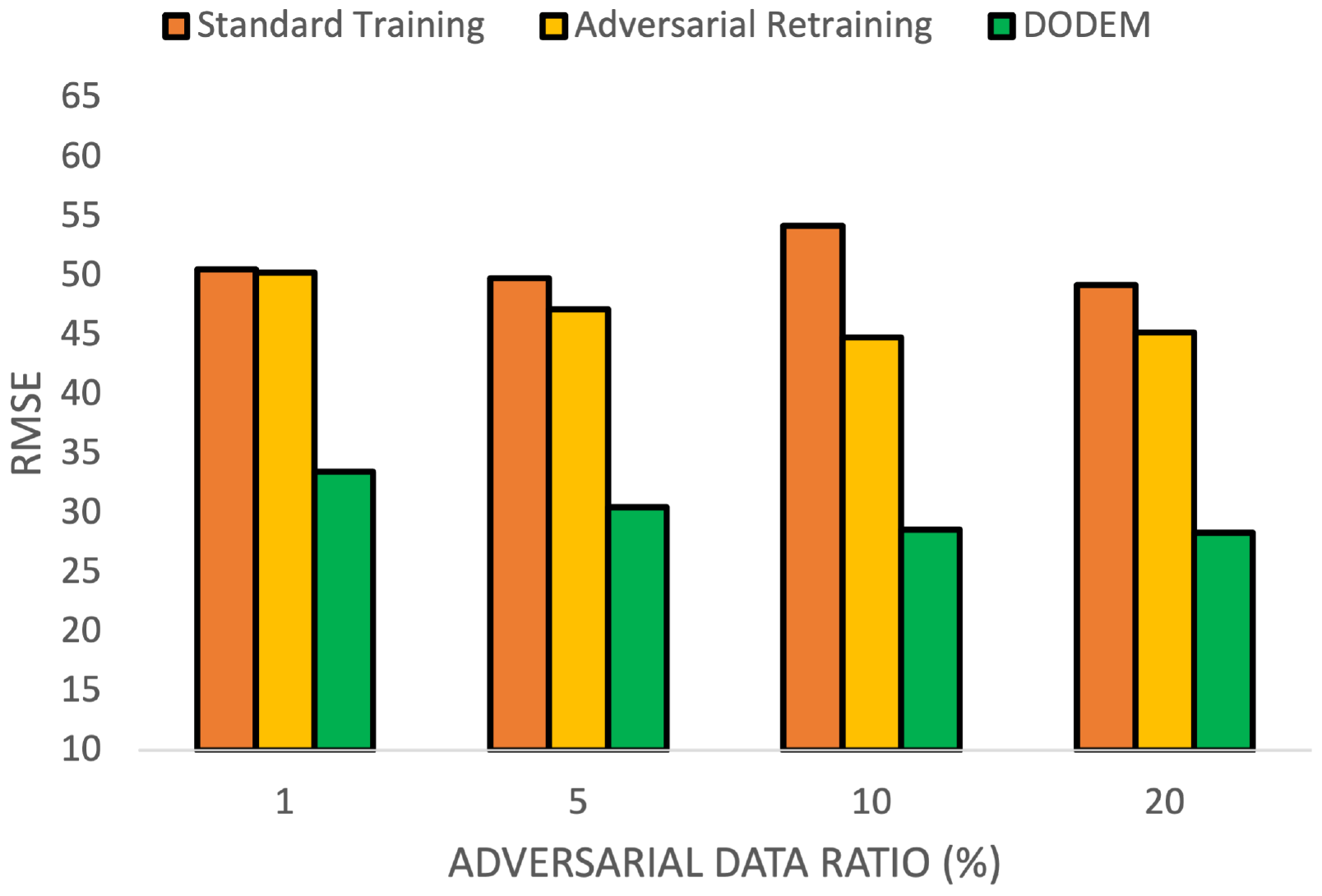}
    \caption{\textbf{FD002}}
\end{subfigure}
\newline
\begin{subfigure}{0.4\linewidth}
    \centering
    \includegraphics[width=\linewidth]{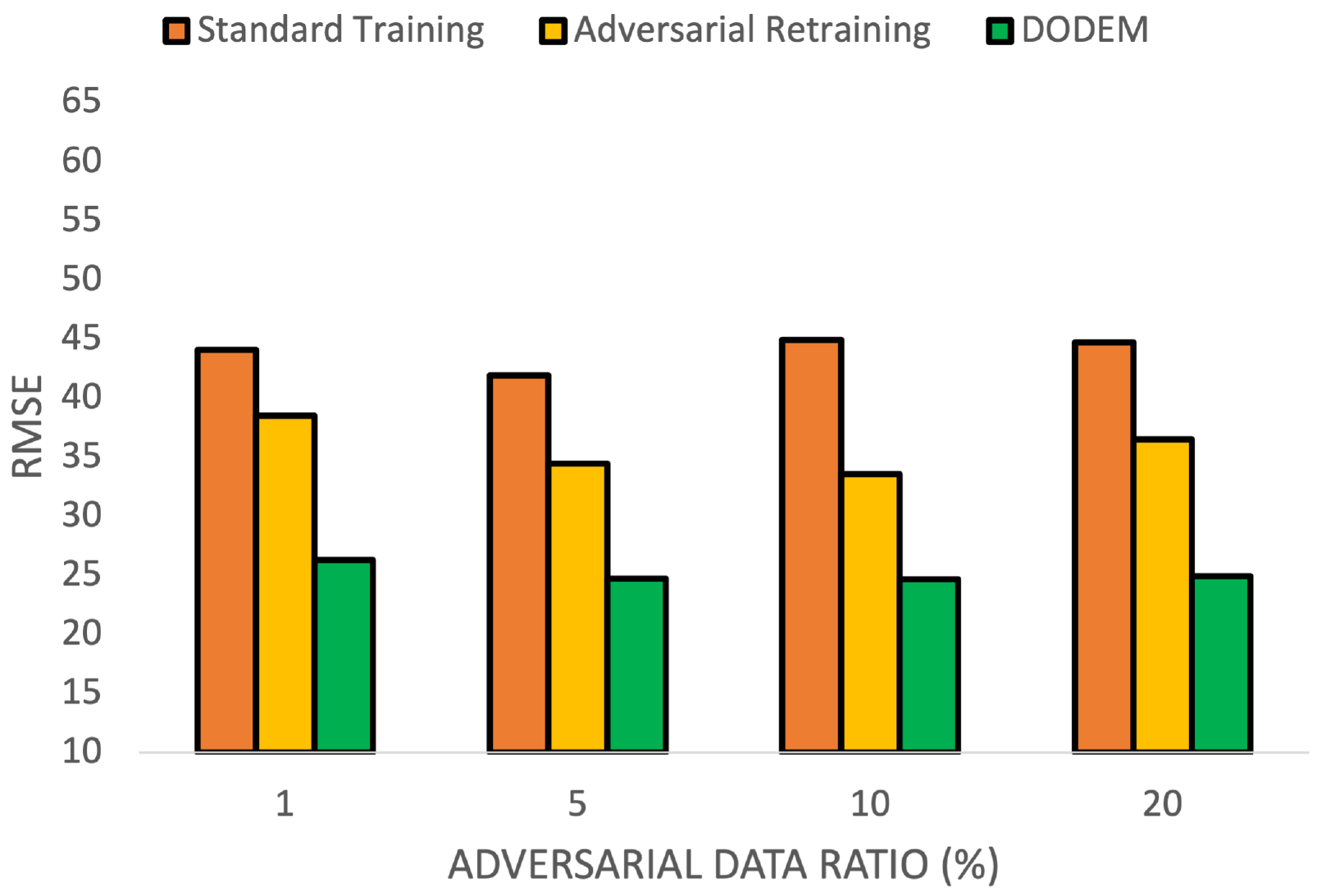}
    \caption{\textbf{FD003}}
\end{subfigure}
\begin{subfigure}{0.4\linewidth}
    \centering
    \includegraphics[width=\linewidth]{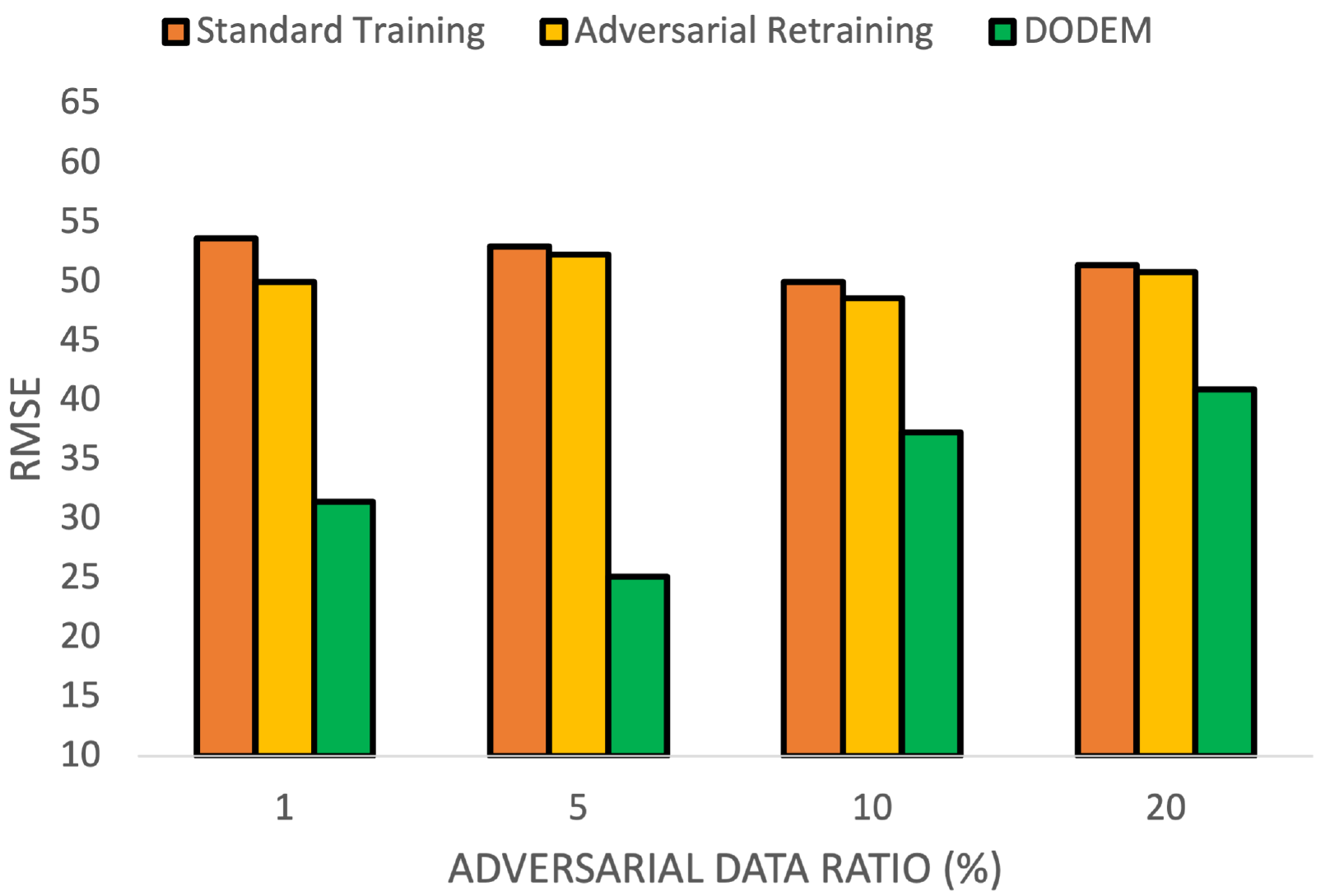}
    \caption{\textbf{FD004}}
\end{subfigure}
\caption{\textit{DODEM} Robustness (FGSM)}
\label{DODEM-robustness-FGSM}
\end{figure}

\begin{figure}
\centering
\begin{subfigure}{0.4\linewidth}
    \centering
    \includegraphics[width=\linewidth]{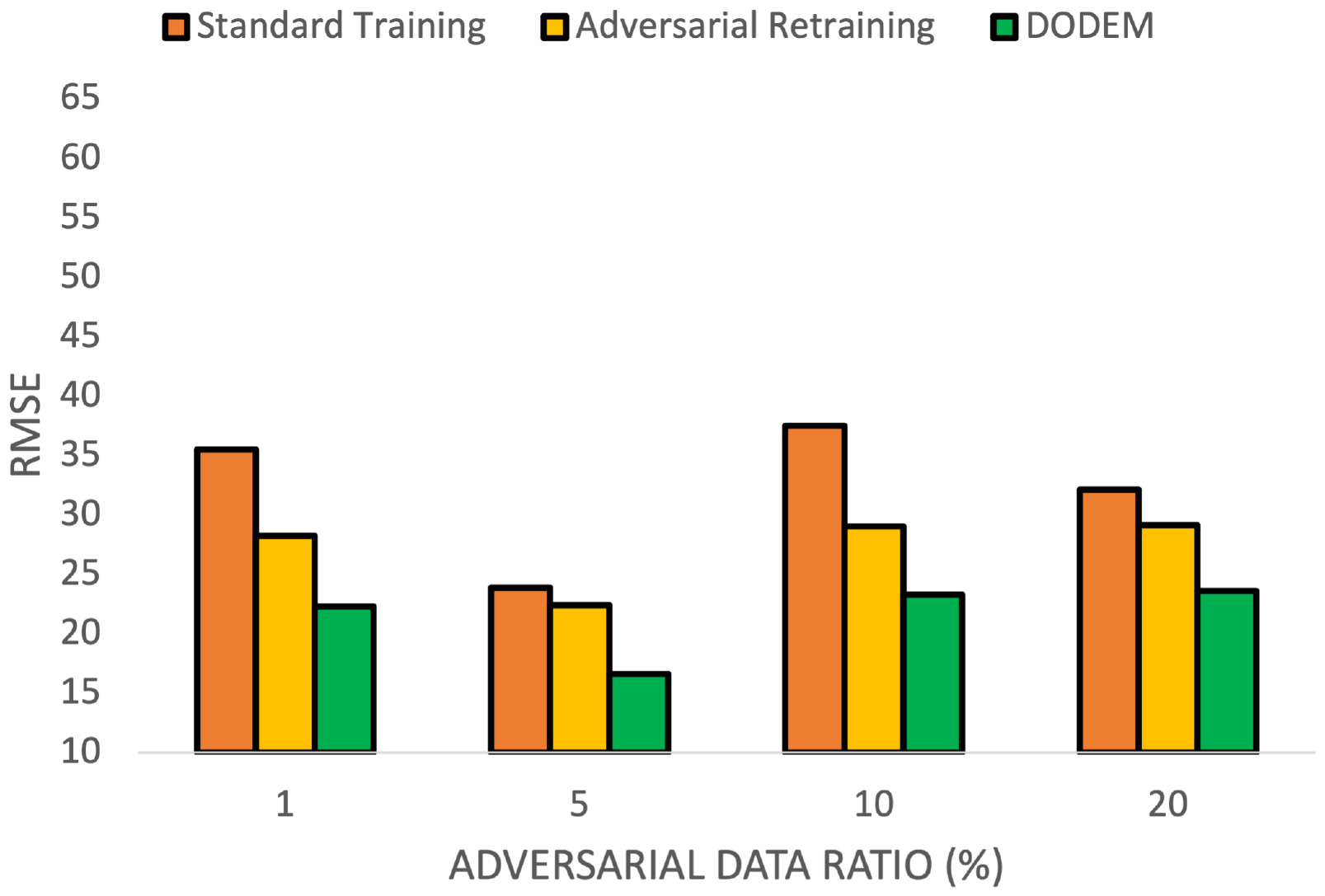}
    \caption{\textbf{FD001}}
\end{subfigure}
\begin{subfigure}{0.4\linewidth}
    \centering
    \includegraphics[width=\linewidth]{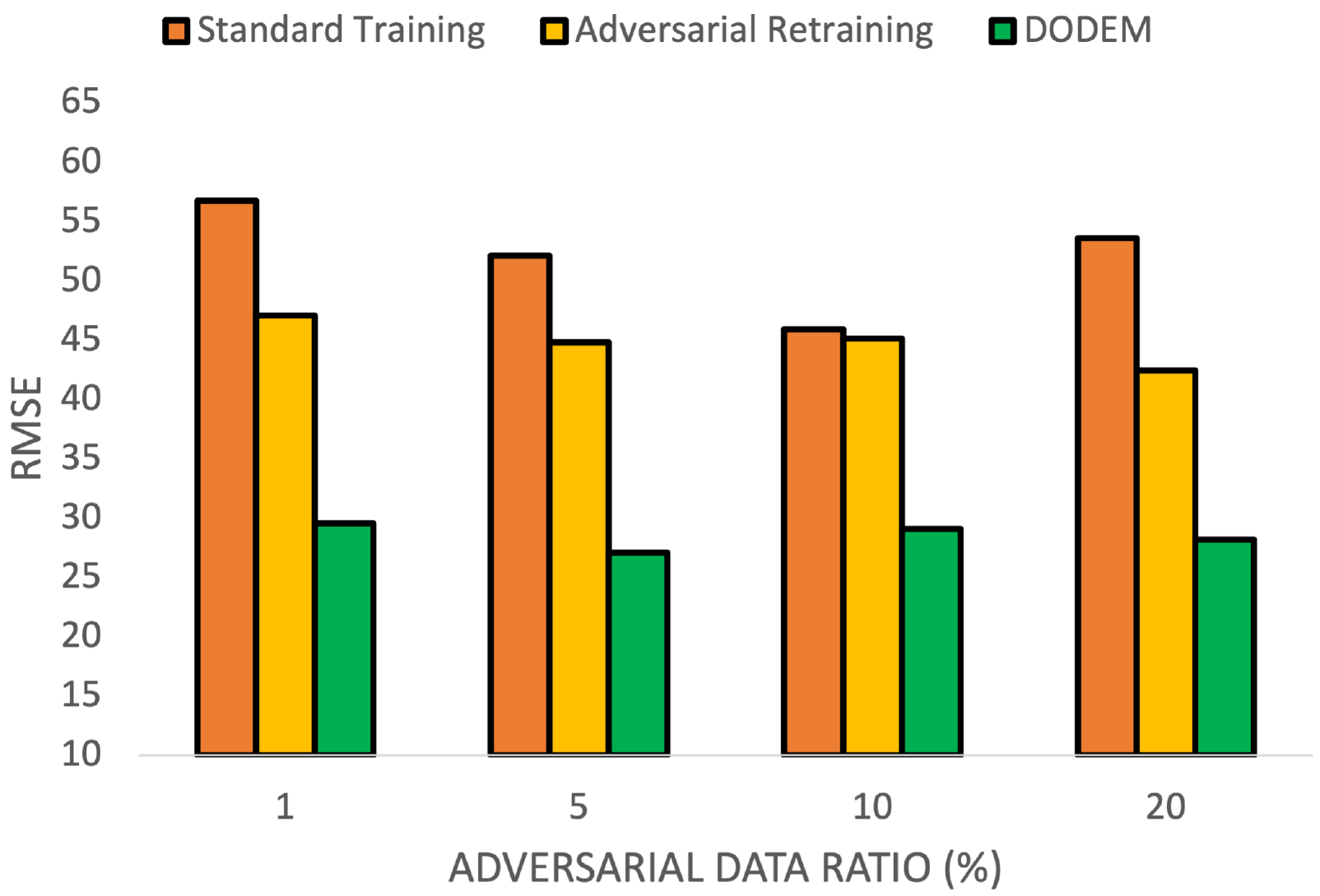}
    \caption{\textbf{FD002}}
\end{subfigure}
\newline
\begin{subfigure}{0.4\linewidth}
    \centering
    \includegraphics[width=\linewidth]{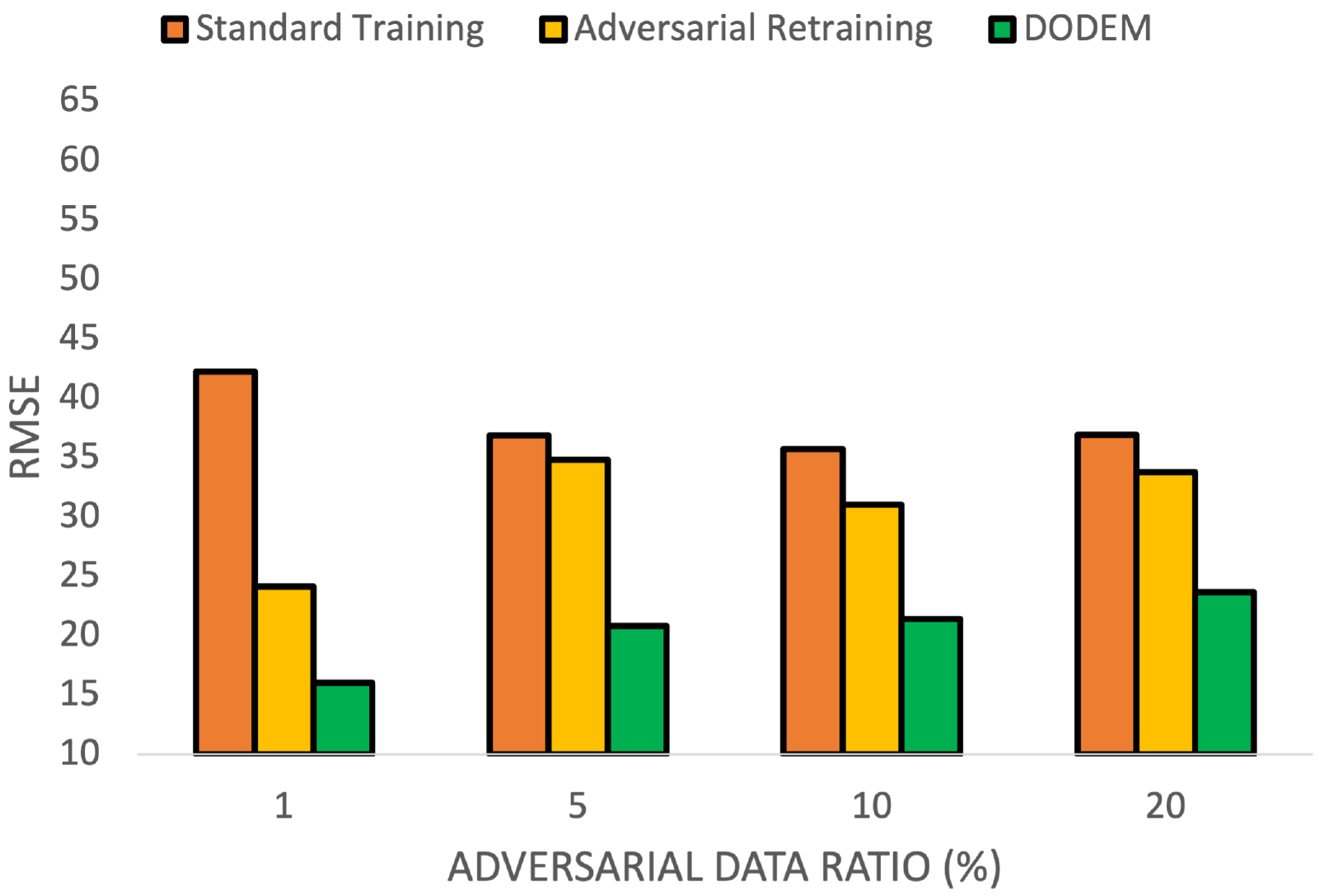}
    \caption{\textbf{FD003}}
\end{subfigure}
\begin{subfigure}{0.4\linewidth}
    \centering
    \includegraphics[width=\linewidth]{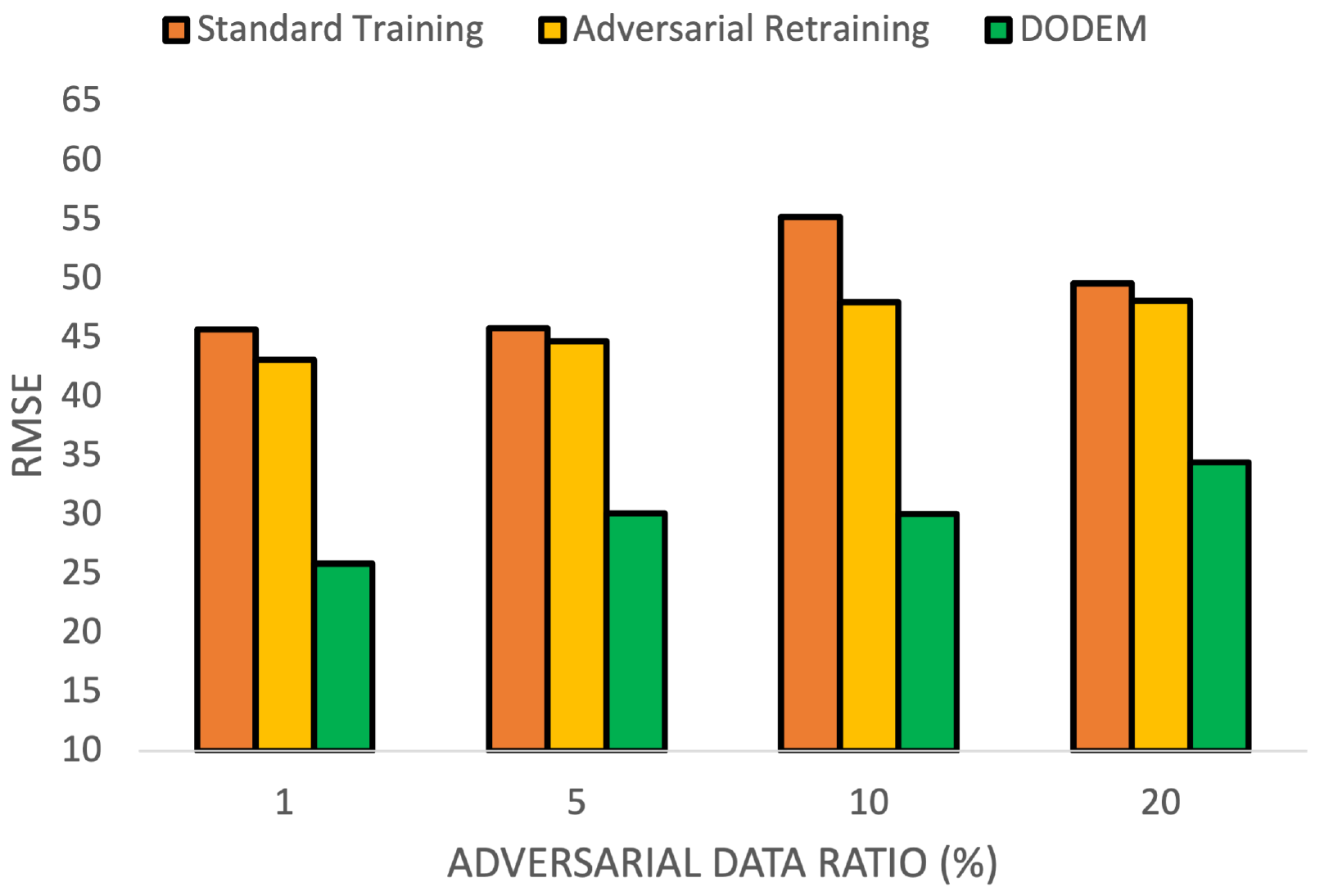}
    \caption{\textbf{FD004}}
\end{subfigure}
\caption{\textit{DODEM} Robustness (BIM)}
\label{DODEM-robustness-BIM}
\end{figure}

\begin{figure}
\centering
\begin{subfigure}{0.4\linewidth}
    \centering
    \includegraphics[width=\linewidth]{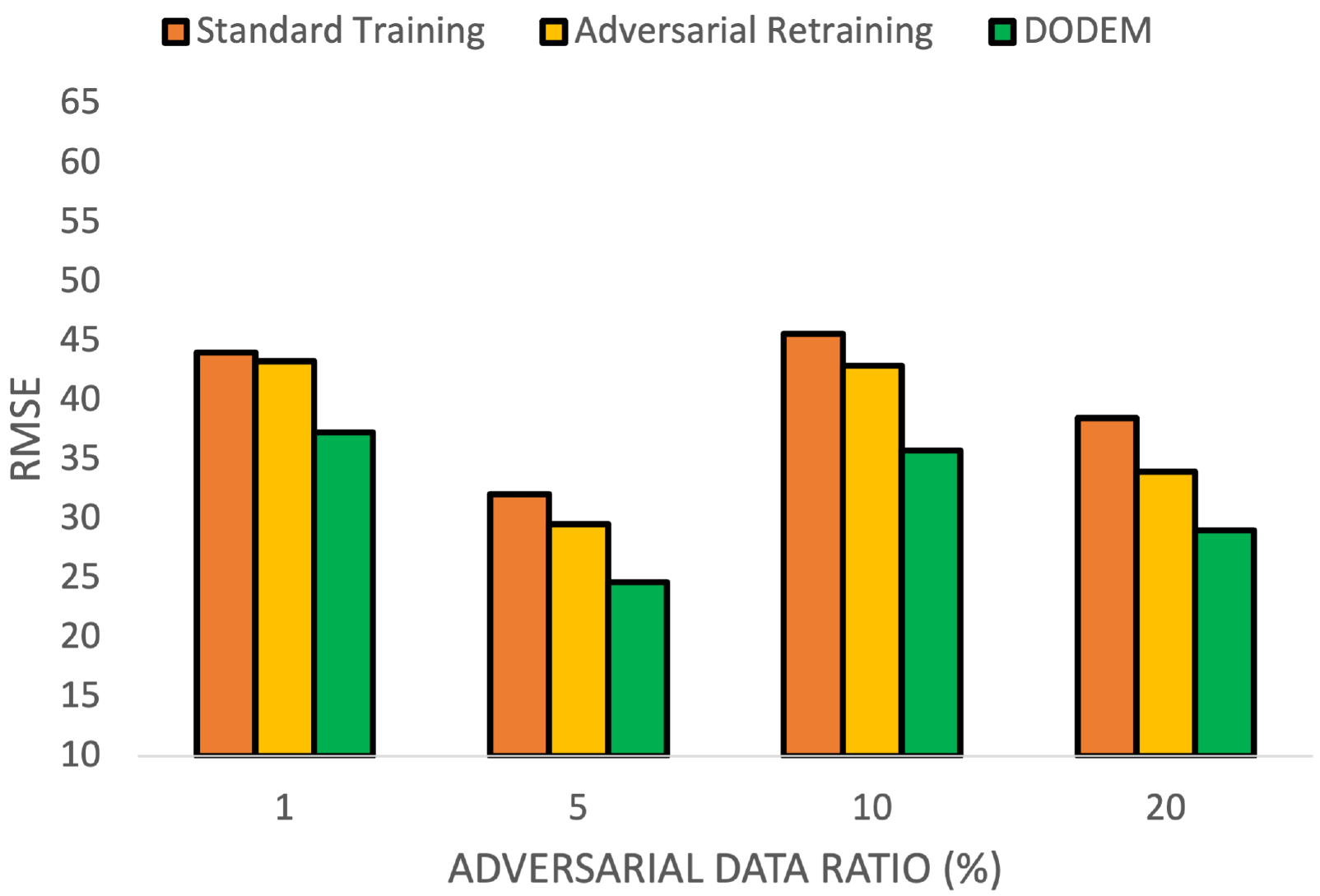}
    \caption{\textbf{FD001}}
\end{subfigure}
\begin{subfigure}{0.4\linewidth}
    \centering
    \includegraphics[width=\linewidth]{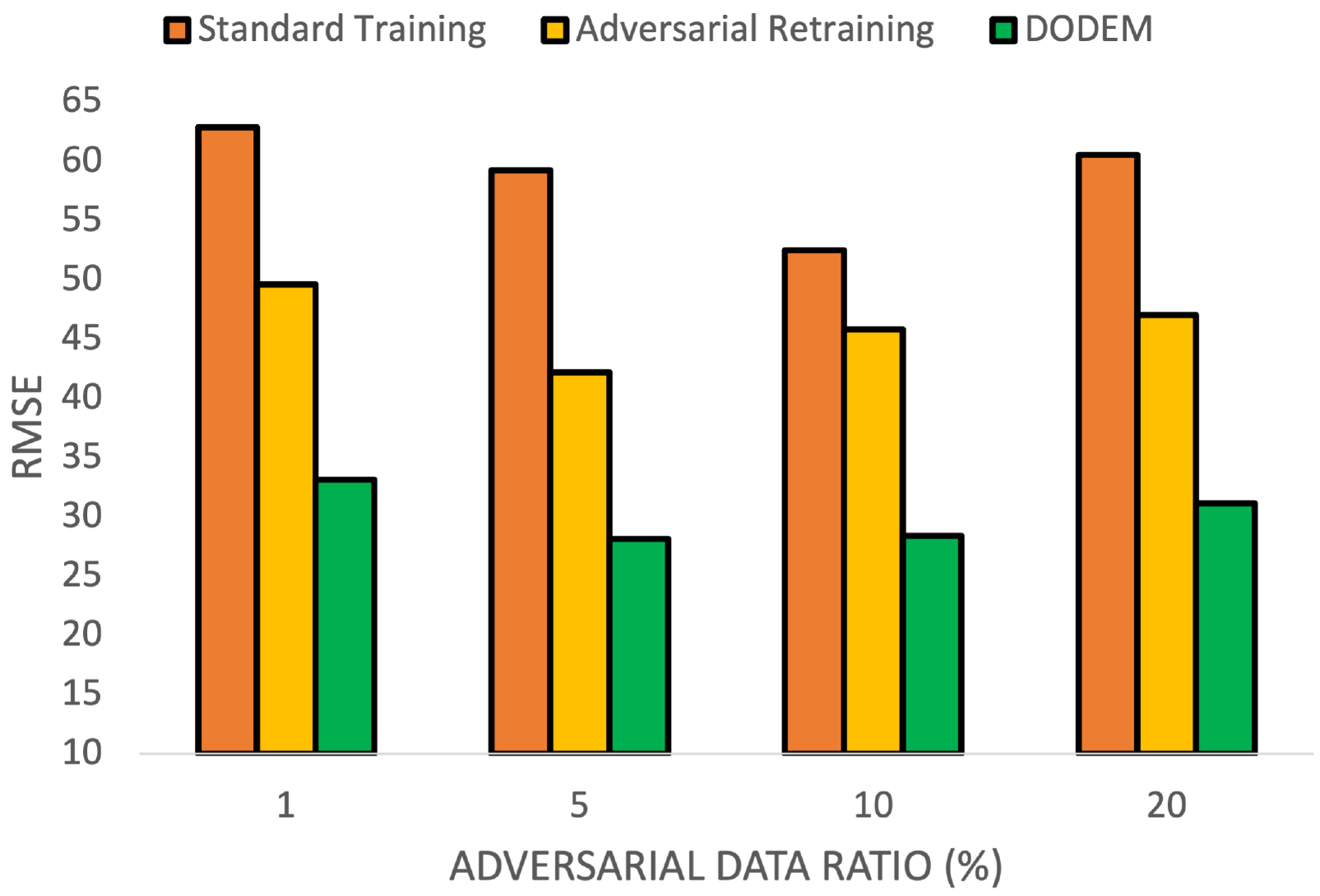}
    \caption{\textbf{FD002}}
\end{subfigure}
\newline
\begin{subfigure}{0.4\linewidth}
    \centering
    \includegraphics[width=\linewidth]{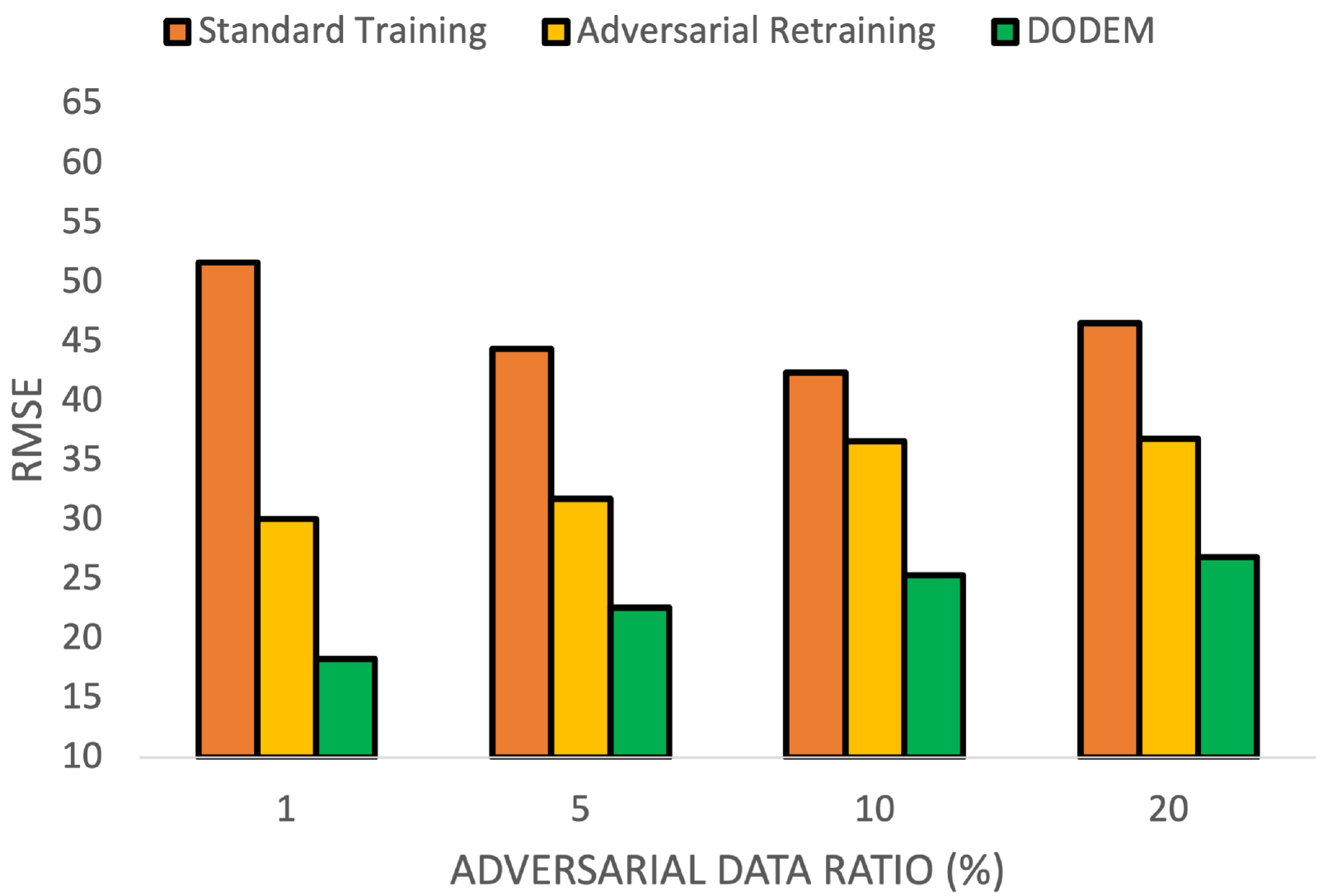}
    \caption{\textbf{FD003}}
\end{subfigure}
\begin{subfigure}{0.4\linewidth}
    \centering
    \includegraphics[width=\linewidth]{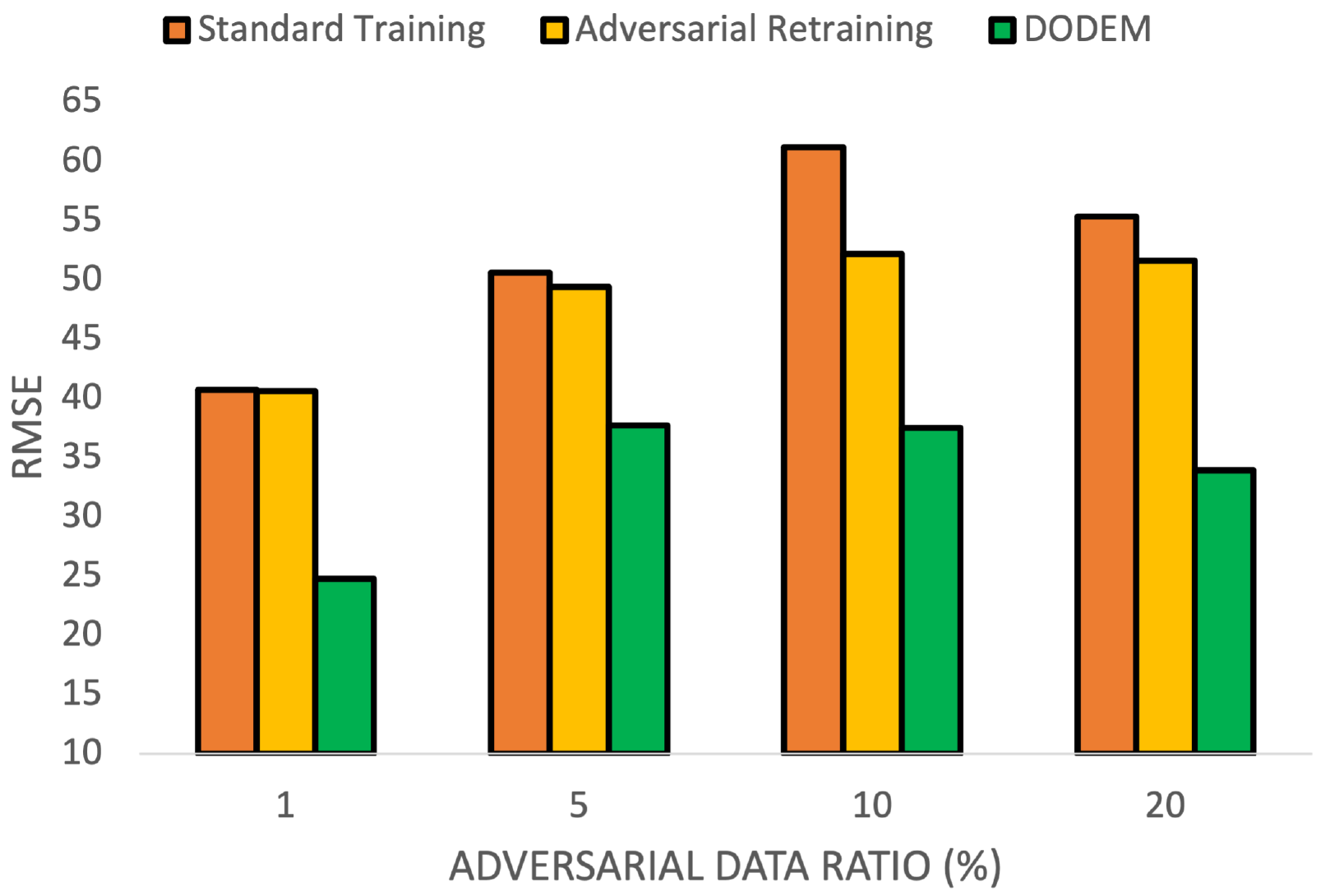}
    \caption{\textbf{FD004}}
\end{subfigure}
\caption{\textit{DODEM} Robustness (MIM)}
\label{DODEM-robustness-MIM}
\end{figure}

\subsection{DODEM Robustness}
\label{proposed-method-robustness}
We compare \textit{DODEM} with two different training settings: standard training and our adversarial retraining with a single model. 
We evaluate \textit{DODEM} with different adversarial test data ratios (ranging from 1\% to 20\%) and attack generation methods (FGSM, BIM, and MIM) to represent a variety of attack scenarios, e.g., a stealth attack with a small number of adversarial samples vs. a heavy attack that has a large number of adversarial samples.
We report the average RMSE for each attack method calculated over perturbed test data which includes both normal and adversarial samples. Figure \ref{DODEM-robustness-FGSM}, Figure \ref{DODEM-robustness-BIM}, and Figure \ref{DODEM-robustness-MIM} present \textit{DODEM} RMSE values under FGSM, BIM, and MIM attacks respectively with standard training and our adversarial retraining setting. Lower RMSE means that a model is more robust against adversarial attacks. Each figure consists of the dataset results. At each sub-figure, x axis denotes the adversarial data ratio in percentage while y-axis is the RMSE. \textit{DODEM} is represented by green color while standard and adversarial training is depicted with orange and yellow colors, respectively. \textit{DODEM} has the smallest RMSE value across all attacks, datasets, and adversarial data ratios consistently. This clearly shows that \textit{DODEM} is the most robust solution against adversarial attacks. We also see that as the dataset complexity increases (from FD001 to FD004), \textit{DODEM} clearly brings more robustness where there is a larger gap among \textit{DODEM} and other methods. This means that our method can perform even better on more complex datasets. We report \textit{DODEM}'s average and maximum robustness improvement over standard training and adversarial retraining in Table \ref{DODEM-improvement-FGSM}, Table \ref{DODEM-improvement-BIM}, and \ref{DODEM-improvement-MIM}. \textit{DODEM} improves the robustness by up to 64.6\% and 52\% for standard and adversarial training respectively. On average, \textit{DODEM} brings up to 49\% and 36.5\% more robustness against adversarial attacks, proving to be an effective defense mechanism against adversarial attacks. 

\begin{table}
\centering
\caption{\textit{DODEM} Robustness Improvement (\%) over Standard Training and Adversarial Retraining (FGSM)}
\scalebox{0.8}{
\begin{tabular}{|c|cc|cc|}
\hline
                                  & \multicolumn{2}{c|}{\textbf{Average}}                         & \multicolumn{2}{c|}{\textbf{Maximum}}                         \\ \hline
\textbf{Dataset/Training Setting} & \multicolumn{1}{c|}{\textbf{Standard}} & \textbf{Adversarial} & \multicolumn{1}{c|}{\textbf{Standard}} & \textbf{Adversarial} \\ \hline
\textbf{FD001}                    & \multicolumn{1}{c|}{26.8}              & 17.7                 & \multicolumn{1}{c|}{44.4}            & 27.3               \\ \hline
\textbf{FD002}                    & \multicolumn{1}{c|}{40.5}              & \textbf{35.5}        & \multicolumn{1}{c|}{47.3}            & 37.3               \\ \hline
\textbf{FD003}                    & \multicolumn{1}{c|}{\textbf{42.7}}     & 29.5                 & \multicolumn{1}{c|}{45.1}            & 31.7              \\ \hline
\textbf{FD004}                    & \multicolumn{1}{c|}{34.9}              & 32.9                 & \multicolumn{1}{c|}{\textbf{52.6}}   & \textbf{52.0}      \\ \hline
\end{tabular}}
\label{DODEM-improvement-FGSM}
\end{table}

\begin{table}
\centering
\caption{\textit{DODEM} Robustness Improvement (\%) over Standard Training and Adversarial Retraining (BIM)}
\scalebox{0.8}{
\begin{tabular}{|c|cc|cc|}
\hline
                                  & \multicolumn{2}{c|}{\textbf{Average}}                         & \multicolumn{2}{c|}{\textbf{Maximum}}                         \\ \hline
\textbf{Dataset/Training Setting} & \multicolumn{1}{c|}{\textbf{Standard}} & \textbf{Adversarial} & \multicolumn{1}{c|}{\textbf{Standard}} & \textbf{Adversarial} \\ \hline
\textbf{FD001}                    & \multicolumn{1}{c|}{33.0}              & 21.4                 & \multicolumn{1}{c|}{37.9}            & 25.8               \\ \hline
\textbf{FD002}                    & \multicolumn{1}{c|}{45.0}              & \textbf{36.5}        & \multicolumn{1}{c|}{48.1}            & 39.6               \\ \hline
\textbf{FD003}                    & \multicolumn{1}{c|}{\textbf{45.4}}     & 33.7                 & \multicolumn{1}{c|}{\textbf{62.0}}            & \textbf{40.3}              \\ \hline
\textbf{FD004}                    & \multicolumn{1}{c|}{38.5}              & 34.7                 & \multicolumn{1}{c|}{45.6}   & 40.0      \\ \hline
\end{tabular}}
\label{DODEM-improvement-BIM}
\end{table}

\begin{table}
\centering
\caption{\textit{DODEM} Robustness Improvement (\%) over Standard Training and Adversarial Retraining (MIM)}
\scalebox{0.8}{
\begin{tabular}{|c|cc|cc|}
\hline
                                  & \multicolumn{2}{c|}{\textbf{Average}}                         & \multicolumn{2}{c|}{\textbf{Maximum}}                         \\ \hline
\textbf{Dataset/Training Setting} & \multicolumn{1}{c|}{\textbf{Standard}} & \textbf{Adversarial} & \multicolumn{1}{c|}{\textbf{Standard}} & \textbf{Adversarial} \\ \hline
\textbf{FD001}                    & \multicolumn{1}{c|}{21.2}              & 15.4                 & \multicolumn{1}{c|}{24.6}            & 16.7               \\ \hline
\textbf{FD002}                    & \multicolumn{1}{c|}{48.6}              & \textbf{34.6}        & \multicolumn{1}{c|}{52.5}            & 38.0               \\ \hline
\textbf{FD003}                    & \multicolumn{1}{c|}{\textbf{49.0}}     & 31.4                 & \multicolumn{1}{c|}{\textbf{64.6}}            & \textbf{39.3}              \\ \hline
\textbf{FD004}                    & \multicolumn{1}{c|}{35.5}              & 31.3                  & \multicolumn{1}{c|}{39.1}   & 39.0      \\ \hline
\end{tabular}}
\label{DODEM-improvement-MIM}
\end{table}

\section{Conclusion}
Industrial Internet of Things (I-IoT), as a typical cyber physical system (CPS), enables fully automated production systems by continuously monitoring devices and analyzing collected data. After an adversary access to the I-IoT system by exploiting its vulnerabilities, they can manipulate legitimate inputs, corrupting ML predictions and causing disruptions in the production systems. Hence, there is a need for sophisticated defense mechanisms that can protect I-IoT systems against adversarial attacks. In this work, we propose double defense mechanism against adversarial attacks. Our first defense method, adversarial attack detection, aims to distinguish adversarial attacks from legitimate ones. To reach that goal, we first extract salient features by using multiple distinct feature extraction methods which are then provided to one-class classifiers. We selected one-class support vector machine (OCSVM) and local outlier factor (LOF). We observed that LOF has a better prediction performance with 89\% average $F_{2}$ score. selective training, our second defense mechanism, decide if an adversarial retraining or standard training should be used based on LOF predictions. If there is an attack, adversarial retraining is used. We modify the standard adversarial retraining by adding adversarial examples via DL methods trained at different datasets, i.e., transfer attack. We also selected different substitute and target models to make the retraining more realistic. Our results show that our adversarial retraining can improve the model robustness by up to 67\% (40\% on average). Most importantly, our double defense mechanism can improve the robustness by up to 64.6\% and 52\% compared to standard and adversarial training respectively under different adversarial attacks.

\bibliographystyle{plain}
\bibliography{references}  






\end{document}